\documentclass[final,numbers,times,3p,12pt]{elsarticle}

\usepackage{amsmath,amssymb,amsfonts,amsthm,eucal, mathrsfs, eufrak}
\usepackage{mathtools}
\usepackage{caption}
\usepackage{graphicx}
\usepackage{subcaption}
\usepackage{url}
\usepackage[usenames,dvipsnames,table,xcdraw]{xcolor}
\usepackage {amssymb}
\usepackage {amsfonts}
\usepackage {mathrsfs}
\usepackage {stmaryrd}
\usepackage {dsfont}

\usepackage {bm}
\usepackage {bbm}
\usepackage {bbold}
\usepackage {dsfont}
\usepackage {yfonts}
\usepackage {bigints}
\usepackage[toc,page]{appendix}
\usepackage {marvosym}
\usepackage {multirow}
\usepackage {hyphenat}
\usepackage {xspace}
\usepackage{lineno}
\usepackage{algorithm,algorithmicx}
\usepackage{algpseudocode}
\usepackage{xcolor}
\usepackage{soul}
\usepackage[tikz]{bclogo}
\usepackage[colorlinks,linkcolor=NavyBlue,citecolor=Red,bookmarksopen,bookmarksnumbered,pdfpagemode=None,pdfstartview=Fit,pdfpagelayout=SinglePage,pdfstartpage=1]{hyperref} 
\biboptions{sort&compress}
\DeclarePairedDelimiter\abs{\lvert}{\rvert}%
\DeclarePairedDelimiter\norm{\lVert}{\rVert}%
\makeatletter
\let\oldabs\abs
\def\abs{\@ifstar{\oldabs}{\oldabs*}}
\let\oldnorm\norm
\def\norm{\@ifstar{\oldnorm}{\oldnorm*}}
\newcommand{\ifun}{\omega\left(\abs{\bm{\xi}} \right)}
\newcommand{\blen}{\abs{\bm{\xi}}}
\newcommand{\blenN}{\abs{\bm{\xi}^N}}
\newcommand{\half}{\frac{1}{2}}
\newcommand{\dvol}{dV_{\bm{x'}}}
\newcounter{Remctr}

\def\ps@pprintTitle{%
\let\@oddhead\@empty
\let\@evenhead\@empty
\def\@oddfoot{}%
\let\@evenfoot\@oddfoot}
\makeatother
\graphicspath{{./figures/}}

\begin{document}

\begin{frontmatter}
	
	\title{A General Numerical Method to Model Anisotropy in Discretized Bond-Based Peridynamics}
	\author{Naveen\ Prakash} \ead{PrakashN2@corning.com} \ead{nprakash@vt.edu}
	\address{1 Science Center Dr., Corning Incorporated, NY 14831}
	
	\begin{abstract}
		This work presents a novel and general method of determining bond micromoduli for anisotropic linear elastic bond-based peridynamics. The problem of finding a discrete distribution of bond micromoduli that reproduces an anisotropic peridynamic stiffness tensor is cast as a least-squares problem. The proposed numerical method is able to find a distribution of bond micromoduli that is able to exactly reproduce a desired anisotropic stiffness tensor provided conditions of Cauchy's relations are met. Examples of all eight possible elastic material symmetries, from triclinic to isotropic are given and discussed in depth. Parametric studies are conducted to demonstrate that the numerical method is robust enough to handle a variety of horizon sizes, neighborhood shapes, influence functions and lattice rotation effects. The proposed method has great potential for modeling of deformation and fracture in anisotropic materials with bond-based peridynamics.
	\end{abstract}
	
	
\end{frontmatter}

\section{Introduction and Motivation}
Peridynamics modeling has been applied to a wide range of problems since it was first introduced by Silling \cite{silling2000}. The advantages of peridynamics, e.g. the ability to use a meshfree type of discretization to solve the equations of motion have allowed the modeling of fracture initiation and propagation with relative ease \cite{silling2005meshfree}. Peridynamics has also been extended to include many multiphysics phenomena - heat conduction \cite{bobaru2010peridynamic, kilic2010peridynamic}, electrical conduction \cite{prakash2016electromechanical, gerstle2008peridynamic}, fluid flow \cite{katiyar2014peridynamic} and corrosion \cite{chen2015peridynamic}. Although there has been a significant amount of work in ordinary\cite{madenci2017ordinary, le2014two, mitchell2015position} and non-ordinary state-based peridynamics\cite{breitenfeld2014quasi, foster2011energy, NME:NME2725, wang2016numerical, silling2007peridynamic, warren2009non, hartmann2020curing, BEHZADINASAB202064}, bond-based peridynamics still attracts a lot of attention in literature. 

One of the areas of interest has been to include anisotropic response within the framework of Peridynamics. For example, Hu and co-workers used bond-based peridynamics to model fracture in uni-directional fiber-reinforced composites using peridynamic bonds of different micromoduli in different directions to introduce anisotropy \cite{hu2012peridynamic, hu2011modeling}. While Hu and co-workers performed a 2D analysis, Oterkus and Madenci \cite{oterkus2012peridynamic} used the same approach for a 3D analysis of composite laminate with multiple plies. Hu et al. used a similar approach to model delamination growth and fatigue in composite structures \cite{hu2016bond, hu2015peridynamic}. Kilic and Madenci on the other hand modeled the fiber and matrix phases of a composite structure and retained the inhomogeneous nature of the material \cite{kilic2009peridynamic}. In more recent work, Mehrmashhadi et al. have presented a stochastic approach to modeling fiber reinforced composites using bond-based peridynamics. Another recent paper by Diana and Casolo used the micropolar version of bond-based peridynamics to model orthotropic materials \cite{diana2019full}. Ghajari et al. has presented more fundamental work in using spherical harmonics expansion to represent the micromodulus as a function of the bond orientation \cite{ghajari2014peridynamic}. A similar approach was followed in Zhou et al. \cite{zhou2017analyzing} as well as in Azdoud et al. \cite{azdoud2013morphing} for orthotropic and transversely isotropic materials. Trageser et al. have presented a detailed analysis of anisotropic bond-based peridynamics models in \cite{trageser2019anisotropic} with a specific focus on 2D plane stress and plane strain models. A similar approach has been taken outside of the peridynamics method as well with granular micromechanics in \cite{chang1990packing, misra2016granular} and virtual internal bond method by Gao and Klein \cite{gao1998numerical}.

Most approaches to model anisotropy so far have used either direct numerical simulations using different micromoduli for different phases to introduce anisotropy in a global sense, or made the micromoduli a function of bond orientation and used a functional form for the micromoduli in terms of the bond orientation to the material's axis to express this relationship. Many of these approaches start with some a priori assumption about the distribution of micromoduli, e.g. all bond micromoduli are constant in the case of isotropy or micromoduli in one direction are stiffer than other directions in case of orthotropy for fiber reinforced composite, for example.

This work presents a very general approach to calibrating micromoduli for \textit{discretized} neighborhoods for anisotropic bond-based models - without any assumptions on the distribution of bond micromoduli a priori. Using a discretized neighborhood or a lattice as a starting point for peridynamic analysis has not been commonly adopted yet. Liu and Hong adopted a discretized bond-based approach in \cite{liu2012discretized} and \cite{liu2012discretizedb} where as Nikravesh and Gerstle presented a state-based peridynamic lattice model (SPLM) using a hexagonal lattice as a starting point \cite{nikravesh2018improved}. Gerstle argues that a discretized peridynamics might be more realistic and computationally efficient \cite{gerstle2015introduction}. Again, discretized lattice based particle methods have existed outside of peridynamics as well, using both local and non-local models e.g. see \cite{buxton2001lattice, schlangen1992simple, schlangen1997fracture, jagota1994spring, curtin1990brittle, ostoja2002lattice, chen2016nonlocal, grah1996brittle}

The least squares method to evaluate bond micromoduli was originally proposed by the author in \cite{prakash2019calibrating}, however only isotropic materials were explored. The current work extends this method to all eight material classes. The main premise of the method is that a collection of bonds in the neighborhood of a peridynamic material particle with a specific distribution of bond stiffnesess or \textit{micromoduli} results in some effective peridynamic stiffness tensor. For example, if all micromoduli are equal, it will result in an isotropic symmetry (note that this is not the only distribution that may lead to isotropic symmetry). The question then is, conversely, can one find a set of micromoduli such that they result in some desired arbitrary stiffness tensor? 

Most published literature deals with specific material classes such as orthotropy, but this work is a first in that it presents a general, robust method for modeling any type of anisotropy in 3D bond-based peridynamic models. No ad-hoc treatment is required for any specific material symmetry. It is shown that if conditions of Cauchy's relations are satisfied by the reference material stiffness tensor, the bond micromoduli can be calibrated such that the effective peridynamic stiffness tensor is exactly equal to the desired reference stiffness tensor regardless of the class of symmetry. 

Restrictions for each of the eight material classes arising from Cauchy's relations and examples are given in section \ref{section_8_symm}. More practical issues such as the effect of horizon size, influence function, shape of the neighborhood and rotation of the neighborhood are also discussed with examples.
	
\section{Peridynamics Preliminaries}
Peridynamics is a continuum theory in which it is assumed that every material particle $\bm{x}$ interacts with every other particle $\bm{x'}$ within some interaction distance called the horizon $\mathcal{H}_{\bm{x}}$  through a \textit{peridynamic bond}. The finite volume surrounding $\bm{x}$ within which these interactions take place is called the neighborhood of the particle. Consider two particles $\bm{x}$ and $\bm{x'}$ in the reference configuration, such that their undeformed relative position is given by $\bm{\xi} = \bm{x'}-\bm{x}$. If $\bm{x}$ and $\bm{x'}$ are displaced by $\bm{u}$ and $\bm{u'}$ respectively, their relative displacement is given by $\bm{\eta} = \bm{u'}-\bm{u}$ such that their relative position in the deformed configuration is given by $\bm{\xi+\eta}$. 

\begin{figure}[H]
	\centering
	\captionsetup{font=footnotesize}
	\captionsetup{justification=centering}
	\includegraphics[width=1\textwidth, trim={3cm 4cm 3cm 4cm},clip]{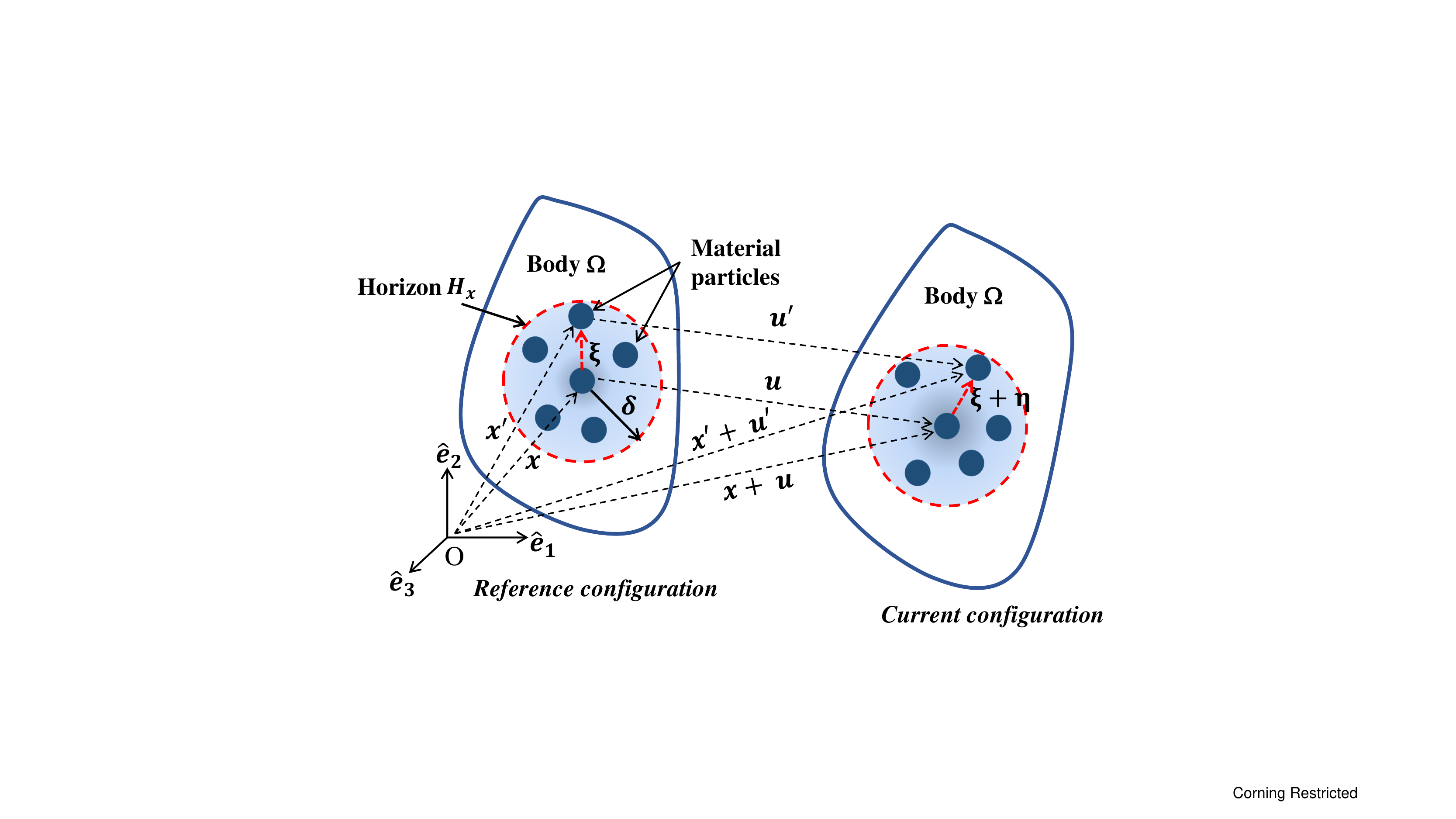}
	\caption{Kinematics of peridynamic material particles, material particle $\bm{x}$ and $\bm{x'}$ correspond to an interacting pair of particles within the horizon region of $\bm{x}$, having a volume $\dvol$ \cite{prakash2017coupled}.}
	\label{PD_schematic}
\end{figure}

\noindent The peridynamic equations of motion for a particle $\bm{x}$ at time t are given by, 
\begin{equation} 
\rho \ddot{\bm{u}}(\bm{x},t) = \int_{\mathcal{H}_{\bm{x}}} \bm{f}( \bm{\eta},\bm{\xi}, t)\; \dvol + \bm{b}(\bm{x},t), \label{Motion_1}
\end{equation}

\noindent where $\rho$ and $\ddot{\bm{u}}$ denote the density and acceleration of the material particle $\bm{x}$, $\bm{f}$ is the pairwise force function (units of force per unit volume squared) of the bond between $\bm{x}$ and $\bm{x'}$ and $\bm{b}$ is the body force per unit volume at $\bm{x}$. The net internal force per unit volume arising from non-local pairwise interactions between particles is obtained from the integral of $\bm{f}$ over the horizon $\mathcal{H}_{\bm{x}}$. Since the formulation does not involve spatial derivatives of displacement, the same governing equations can be applied in the presence of discontinuities.

According to Silling \cite{silling2000}, a micro-elastic material is defined as one in which the internal force is the gradient of a scalar potential \textit{w}$( \bm{\eta},\bm{\xi})$ i.e., 
\begin{equation} 
\bm{f}(\bm{\eta},\bm{\xi},t) = \frac{\partial w( \bm{\eta},\bm{\xi})} {\partial \bm{\eta}}. \label{Energy_1}
\end{equation} 

For an elastic material, the scalar potential refers to a micro-elastic strain energy, i.e. strain energy density per unit volume stored in the material. The material is said to be linear micro-elastic \cite{silling2000} if \textit{w} is chosen to be,
\begin{equation} 
w(\bm{\xi,\eta}) = \frac{c(\bm{\xi})s^2(\bm{\eta},\bm{\xi})\blen}{2}, \label{Energy_2}
\end{equation} 

\noindent where $c(\bm{\xi})$ is a constant called the micromodulus, $\blen$ is the magnitude of the reference bond vector $\bm{\xi}$ and s is the stretch of the bond which is given by, 
\begin{equation} 
s = \frac{|\bm{\xi + \eta}|-\blen}{\blen}. \label{Stretch}
\end{equation} 

The micromodulus $c(\bm{\xi})$ can depend on the reference bond vector $\bm{\xi}$, however for homogeneous isotropic materials the micromodulus is generally assumed to be independent of the direction of the bond. For the current purposes, this dependence is retained. If Eq. \eqref{Energy_2} is substituted in Eq. \eqref{Energy_1}, the expression for the internal force density in a peridynamic bond for a homogeneous linear isotropic material can be derived such that the internal force $\bm{f}$ is given by,
\begin{equation} 
\bm{f\left(\eta, \xi \right)} = c(\bm{\xi}) s\frac{\bm{\xi+\eta}}{|\bm{\xi+\eta}|}, \label{Force_0}\end{equation} 

\noindent where the force in the bond is a function of the stretch of the bond $s$, the micromodulus $c(\bm{\xi})$, and acts in the direction of the deformed bond as indicated by the unit vector $\left(\bm{\xi+\eta} \right)/ \left(|\bm{\xi+\eta}| \right) $. This is commonly known as the \textit{linear microelastic} model of peridynamics and the reader is referred to Silling\cite{silling2000, silling2010peridynamic} for a detailed derivation of $\bm{f}$ and Eq. \eqref{Force_0}. In this general form, bond stretch is a nonlinear function of $\bm{u'-u}$, the relative displacement of the bond and hence the force - displacement relationship is nonlinear. 

This constitutive model can be linearized by assuming small deformations, i.e. $|\bm{\eta}| \ll 1 $. Rewriting Eq. \eqref{Force_0},
\begin{equation} \bm{f\left(\eta, \xi \right)} = c(\bm{\xi}) \left(\frac{|\bm{\xi + \eta}|-|\bm{\xi}|}{|\bm{\xi}|}\right)\bm{\frac{\xi+\eta}{|\xi+\eta|}}, \label{Force_1}
\end{equation} 

\begin{equation} \Rightarrow \bm{f\left(\eta, \xi \right)} = c(\bm{\xi}) \left(\frac{1}{|\bm{\xi}|} - \frac{1}{|\bm{\xi+\eta}|} \right)\bm{\xi+\eta} \label{Force_2}
\end{equation} 

\noindent For $|\bm{\eta}| \ll 1 $, using Taylor series expansion about $\bm{\eta = 0}$ and ignoring higher order terms of $\bm{\eta}$,
\begin{equation} \bm{f\left(\eta, \xi \right)} = \bm{f\left(0, \xi \right)} + \frac{\partial \bm{f}}{\partial \bm{\eta}}\bigg|_{\substack{\bm{\eta = 0}}} \bm{\eta}, \label {Force_3}
\end{equation} 

\noindent where the first term goes to zero, meaning that the internal force in the bond is zero in the undeformed configuration. Substituting for $\bm{f}$ in the second term of Eq. \eqref{Force_3} from Eq. \eqref{Force_2},
\begin{equation}
\bm{f\left(\eta, \xi \right)} =  \left[ c(\bm{\xi}) \frac{\partial}{\partial \bm{\eta}} \left( \frac{1}{|\bm{\xi}|} - \frac{1}{|\bm{\xi + \eta}|} \right) \otimes \left( \bm{\xi + \eta} \right) \right] \bigg|_{\substack{\bm{\eta = 0}}} \bm{\eta}+ c(\bm{\xi}) \left[  \left( \frac{1}{|\bm{\xi}|} - \frac{1}{|\bm{\xi + \eta}|} \right) \frac{\partial \left( \bm{\xi + \eta}\right)}{\partial \bm{\eta}} \right] \bigg|_{\substack{\bm{\eta = 0}}} \bm{\eta}. \label{Force_4}
\end{equation}

\noindent The second term in Eq. \eqref{Force_4} when evaluated at $\bm{\eta = 0}$ is zero and Eq. \eqref{Force_4} becomes,
\begin{equation}
\bm{f\left(\eta, \xi \right)} = \left[ c(\bm{\xi}) \frac{\partial}{\partial \bm{\eta}} \left( - \frac{1}{|\bm{\xi + \eta}|} \right) \otimes \left( \bm{\xi + \eta} \right) \right] \bigg|_{\substack{\bm{\eta = 0}}} \bm{\eta}, \label{Force_5}
\end{equation}

\noindent where the partial derivative of $ 1/|\bm{\xi}| $ with respect to $\bm{\eta}$ is zero. Evaluating Eq. \eqref{Force_5} and substituting $\bm{\eta = 0}$,
\begin{equation}
\bm{f\left(\eta, \xi \right)} = \left[ c(\bm{\xi}) \frac{\bm{\xi} \otimes \bm{\xi} }{|\bm{\xi}|^3} \right] \bm{\eta}, \label{Force_6}
\end{equation}

\noindent where the quantity in brackets is a second order tensor with components that are a function of the micromodulus $c(\bm{\xi})$ and $\bm{\xi}$. Rewriting Eq. \eqref{Force_6},
\begin{equation}
\bm{f\left(\eta, \xi \right)} = \left[ c(\bm{\xi}) \frac{\bm{\xi}}{|\bm{\xi}|} \otimes \frac{\bm{\xi}}{|\bm{\xi}|} \right] \frac{\bm{\eta}}{|\bm{\xi}|}, \label{Force_7}
\end{equation}

\noindent and using the tensor product rule $\left(\bm{a} \otimes \bm{b} \right)\bm{c} = \bm{\left(b.c\right)a}$, Eq. \eqref{Force_7} can be written as,
\begin{equation}
\bm{f\left(\eta, \xi \right)} =  c(\bm{\xi}) \frac{\bm{\xi}}{|\bm{\xi}|} \cdot \frac{\bm{\eta}}{|\bm{\xi}|} \frac{\bm{\xi}}{|\bm{\xi}|}. \label{Force_8}
\end{equation}

Recognizing that $\bm{\xi}/|\bm{\xi}|$ is nothing but the unit vector in the undeformed configuration of the bond and that $\bm{\eta} \cdot \bm{\xi}/|\bm{\xi}| $ is the component of relative displacement $\eta$ in the direction of the undeformed bond, Eq. \eqref{Force_8} can be written as ,
\begin{equation}
\bm{f\left(\eta, \xi \right)} =  c(\bm{\xi}) \frac{\eta_n}{\blen} \frac{\bm{\xi}}{\blen},  \label{Force_9}
\end{equation}

\noindent where $\eta_n$ denotes the component of relative displacement in the direction of the undeformed bond. Comparing Eq. \eqref{Force_9} to Eq. \eqref{Stretch}, it is found that the term $\eta_n/\blen $ is the linearized stretch $s$ and the direction of the force is changed from the deformed bond direction to the undeformed bond direction. Substituting for the internal force density from Eq. \eqref{Force_9} in Eq. \eqref{Motion_1}, the peridynamic linear momentum equation with a linearized force - displacement relationship becomes,
\begin{equation} \rho \ddot{\bm{u}}(\bm{x},t) = \int_{\mathcal{H}_{\bm{x}}} c(\bm{\xi}) \frac{\eta_n}{\blen} \frac{\bm{\xi}}{\blen}  \dvol + \bm{b}(\bm{x},t).\label{Motion_4}
\end{equation}

The corresponding strain energy density for the linearized version of the linear microelastic model is given by,
\begin{equation} 
w(\bm{\xi,\eta}) = \half c(\bm{\xi}) \frac{ \eta_{n}^{2}}{\blen} \label{Energy_5}
\end{equation} 

The strain energy density stored at a material point can be written as the integral over all bonds connected to the material point,
\begin{equation}
W^{PD}(\bm{x}) = \half\int_{\mathcal{H}_{\bm{x}}} \half c(\bm{\xi}) \frac{ \eta_{n}^{2}}{\blen}  \dvol, \label{Energy_6}
\end{equation}

\noindent where the $1/2$ outside the integral appears to account for the fact that the material point is assumed to stored half the strain energy stored in each bond. 

Note that the material model is still general with respect to material symmetry. One way to ensure isotropy is to assume that the micromodulus function is constant i.e. $c(\bm{\xi}) = c$. Then the micromodulus $c$ in Eq. \eqref{Motion_4} can then be evaluated by comparing the stored peridynamic strain energy density to that obtained using classical continuum mechanics principles. The expression for the micromodulus $c$ for isotropy using this assumption can be derived in 3D as \cite{silling2005meshfree}, 
\begin{equation}
c^{3D} = \frac{18K}{\pi \delta^4},\label{silling3D}
\end{equation}

\noindent where, $K$ is the bulk modulus of the material. Similar expressions for the micromodulus can be derived for 2D plane stress and plane strain conditions as well. 

The general linearized bond-based model, with the strain energy density as shown in Eq. \eqref{Energy_6} is taken as the starting point for the proposed numerical method going forward and is used to derive the effective peridynamic stiffness tensor as shown in the next section.

\section{Least-Squares Method to Calibrating Micromoduli} \label{opti}

Assume that under small strains and displacements, the macroelastic strain energy density under a homogeneous strain field for linear elastic material is given by,
\begin{equation}
W^{PD}(\bm{x}) = \half\int_{\mathcal{H}_{\bm{x}}} \half \ifun c\left(\bm{\xi}\right) \frac{ \eta_{n}^{2}}{\blen}  \dvol, \label{Cijkl_1}
\end{equation}

\noindent which is the same as Eq. \eqref{Energy_6} but with an additional influence function $\ifun$ that weights contributions of individual bonds to the strain energy stored based on their undeformed bond lengths. Note that this equation can be derived similarly starting with the inclusion of the influence function in Eq. \eqref{Energy_2}, however is not done to keep the text succinct. 

Assume that the relative displacements of the bonds $\eta_{n}$ is caused by some imposed homogeneous strain field $\bm{\varepsilon}$. Taking the derivative of Eq. \eqref{Cijkl_1} with respect to the second order strain tensor,
\begin{equation}
\frac{\partial W^{PD}}{\partial \varepsilon_{kl}} 
= \half\int_{\mathcal{H}_{\bm{x}}} \ifun c\left(\bm{\xi}\right) \frac{ \eta_{n}}{\blen}
\frac{\partial \eta_n}{\partial \varepsilon_{kl}}  \dvol, \label{Cijkl_2}
\end{equation}

\noindent where the scalar relative displacement can be written in terms of the bond length and the strain field as, 
\begin{equation}
\eta_{n} = \frac{\xi_i \varepsilon_{ij} \xi_j}{\blen}, \label{rel_disp_strain}
\end{equation}

\noindent and the derivative can be written as, 
\begin{equation}
\frac{\partial \eta_n}{\partial \varepsilon_{kl}}
= \frac{\xi_k \xi_l}{\blen}. \label{der_rel_disp_strain}
\end{equation}

\noindent Substituting into Eq. \eqref{Cijkl_2} and taking the derivative again,
\begin{equation}
\frac{\partial W^{PD}}{\partial \varepsilon_{ij} \partial \varepsilon_{kl}} 
= \half \int_{H_{\bm{x}}} \ifun c\left(\bm{\xi}\right)\; \frac{\xi_i \xi_j \xi_k \xi_l}{\blen^3 }
\dvol. \label{Cijkl_3}
\end{equation}

\noindent which is nothing but the fourth order stiffness tensor,
\begin{equation}
\mathbb{C}^{PD}_{ijkl} = \half \int_{H_{\bm{x}}} \ifun c\left(\bm{\xi}\right)\; \frac{\xi_i \xi_j \xi_k \xi_l}{ \blen^3 } \dvol, \label{Cijkl_4}
\end{equation} 

\noindent or in vector notation can be written as,
\begin{equation}
\mathcal{\bf{C}}^{PD} = \half \int_{H_{\bm{x}}} \ifun c\left(\bm{\xi}\right) \; \frac{\bm{\xi} \otimes \bm{\xi} \otimes \bm{\xi} \otimes \bm{\xi}}{ \blen^3 } \dvol, \label{Cijkl_5} 
\end{equation} 

\noindent where the integral is over the horizon of particle $\bm{x}$ which covers an infinite number of it's neighbors $\bm{x'}$ and the range of indices $ijkl$ are from 1 to 3. Observe from the form of Eq. \eqref{Cijkl_4} and Eq. \eqref{Cijkl_5} that $\mathcal{\bf{C}}^{PD}$ has the following symmetries commonly known as the minor and major symmetries respectively which are also present in the classical form of the fourth order material stiffness tensor,
\begin{equation}
\mathbb{C}^{PD}_{ijkl} = \mathbb{C}^{PD}_{jikl} = \mathbb{C}^{PD}_{ijlk}, \label{major}
\end{equation}

\noindent and, 
\begin{equation}
\mathbb{C}^{PD}_{ijkl} = \mathbb{C}^{PD}_{klij}. \label{minor}
\end{equation}

\noindent In addition, note that the peridynamics stiffness tensor also contains an \textit{additional} symmetry of the inner indices $j$ and $k$, also known as Cauchy's relations or Cauchy's symmetry \cite{love2013treatise, clayton2010nonlinear, trageser2019bond} \footnote{We refer to this as Cauchy's relations henceforth},
\begin{equation}
\mathbb{C}^{PD}_{ijkl} = \mathbb{C}^{PD}_{ikjl}, \label{cauchy}
\end{equation}

\noindent thus making the stiffness tensor fully symmetric. In other words, any two indices in the peridynamic stiffness tensor may be interchanged having no effect on the stiffness tensor. There are no other restrictions on the peridynamic stiffness tensor other than those given in Eq. \eqref{major}, Eq. \eqref{minor} and Eq. \eqref{cauchy}, i.e. any material, not necessarily just isotropic can be represented using Eq. \eqref{Cijkl_5} provided the necessary symmetry conditions are met.

The discrete version for a discretized neighborhood\footnote{The collection of bonds here is referred to as the `neighborhood', some works also refer to this as the `family' of the particle.} at some particle in the domain can be written by reducing the integral to a finite summation following the method of Silling and Askari \cite{silling2005meshfree} in index notation as,
\begin{equation}
\mathbb{C}^{PD}_{ijkl} = \half \sum_{N=1}^{M} \omega^N c^N\; \frac{\xi^N_i \xi^N_j \xi^N_k \xi^N_l}{ \blenN^3 } \Delta V_N. \label{discrete_c_peri}
\end{equation} 

\noindent This expression is written for a particle $\bm{x}$ as a summation over it's Nth neighbor where N varies from 1 to M where M is the total number of neighbors, and therefore the total number of bonds. For example, the stiffness term $C^{PD}_{1111}$ can be written as, 
\begin{equation}
\mathbb{C}^{PD}_{1111} = \half \sum_{N=1}^{M}  \omega^N c^N\; \frac{\xi^N_1 \xi^N_1 \xi^N_1 \xi^N_1}{ \blenN^3  } \Delta V_N.
\end{equation}

\noindent Note that quantities for each bond appearing in Eq. \eqref{discrete_c_peri} - e.g. $\omega^N, c^N$ etc. are identified with a superscript $N$ except for the volume which is written with an $N$ in the subscript due to convention. Voigt notation can be used to simplify, where by taking advantage of minor and major symmetries, the fourth order stiffness tensor can be written as a 6 $\times$ 6 matrix. Therefore Eq. \eqref{discrete_c_peri} can be written in terms of the Voigt indices as,
\begin{equation}
\tilde{C}^{PD}_{\alpha\beta} = \half \sum_{N=1}^{M} \omega^N c^N \; \frac{ \zeta^N_{\alpha} \zeta^N_{\beta} }{ \blenN^3  } \Delta V_N, \label{discrete_c_peri_2}
\end{equation} 

\noindent where $\alpha$ and $\beta$ are the two Voigt indices that vary from 1 to 6 such that they relate to cartesian indices $ijkl$ as: 1$\rightarrow$11, 2$\rightarrow$22, 3$\rightarrow$33, 4$\rightarrow$23, 5$\rightarrow$31 and 6$\rightarrow$12. For example $C_{56}$ in Voigt notation corresponds to $C_{3112}$ in conventional notation, or $C_{1312}$ due to minor symmetry. Similarly, $\zeta_1 = \xi_1 \xi_1$, $\zeta_2 = \xi_2 \xi_2$, $\zeta_1 = \xi_3 \xi_3$, $\zeta_4 = \xi_2 \xi_3$, $\zeta_5 = \xi_1 \xi_3$ and $\zeta_6 = \xi_1 \xi_2$.

For each one of the 21 independent constants, (\ref{discrete_c_peri_2}) can be written as a dot product of two vectors by separating out the bond micromoduli, 
\begin{equation}
\tilde{C}^{PD}_{\alpha\beta} = \left[\sum_{N=1}^{M} \half \; \omega^N \; \frac{ \zeta^N_{\alpha} \zeta^N_{\beta} }{ \blenN^3  } \Delta V_N \right] .c^N, \label{discrete_c_peri_3}
\end{equation} 

\noindent where the length of the vectors are equal to the number of neighbors or the number of bond micromoduli to be evaluated - M. For example, for the 11 component,
\begin{equation}
\begin{bmatrix} 
\tilde{C}^{PD}_{11} 
\end{bmatrix} = \\ \\ 
\begin{bmatrix} 
\half\omega^1 \; \frac{ \zeta^1_{1} \zeta^1_{1} }{ \abs{\bm{\xi}^1}^3 } \Delta V_1 &  
\half\omega^2 \; \frac{ \zeta^2_{1} \zeta^2_{1} }{ \abs{\bm{\xi}^2}^3 } \Delta V_2	& 
\half\omega^3 \; \frac{ \zeta^3_{1} \zeta^3_{1} }{ \abs{\bm{\xi}^3}^3 } \Delta V_3 & \cdots &
\half\omega^M \; \frac{ \zeta^M_{1} \zeta^M_{1} }{ \abs{\bm{\xi}^M}^3 } \Delta V_M\\ 
\end{bmatrix}
\begin{bmatrix} 
c^1\\ 
c^2\\
c^3\\
\vdots\\
c^M 
\end{bmatrix}.
\end{equation}

\noindent If the 21 independent stiffness components $\tilde{C}^{PD}_{\alpha\beta}$ are arranged as components of a vector $[\tilde{C}^{PD}_{11} \; \tilde{C}^{PD}_{12} \; \tilde{C}^{PD}_{13}\; ... \;\tilde{C}^{PD}_{66}]^T$ then, 
\begin{equation}
\begin{bmatrix} 
\tilde{C}^{PD}_{11}\\ \\
\tilde{C}^{PD}_{12}\\ \\
\tilde{C}^{PD}_{13}\\
\vdots\\
\tilde{C}^{PD}_{66}
\end{bmatrix} = 
\begin{bmatrix}

\half\omega^1 \; \frac{ \zeta^1_{1} \zeta^1_{1} }{ \abs{\bm{\xi}^1}^3 } \Delta V_1 &  
\half\omega^2 \; \frac{ \zeta^2_{1} \zeta^2_{1} }{ \abs{\bm{\xi}^2}^3 } \Delta V_2	& 
\half\omega^3 \; \frac{ \zeta^3_{1} \zeta^3_{1} }{ \abs{\bm{\xi}^3}^3 } \Delta V_3 & \cdots & 
\half\omega^M \; \frac{ \zeta^M_{1} \zeta^M_{1} }{ \abs{\bm{\xi}^M}^3 } \Delta V_M\\ \\

\half\omega^1 \; \frac{ \zeta^1_{1} \zeta^1_{2} }{ \abs{\bm{\xi}^1}^3 } \Delta V_1 &  
\half\omega^2 \; \frac{ \zeta^2_{1} \zeta^2_{2} }{ \abs{\bm{\xi}^2}^3 } \Delta V_2	& 
\half\omega^3 \; \frac{ \zeta^3_{1} \zeta^3_{2} }{ \abs{\bm{\xi}^3}^3 } \Delta V_3 & \cdots & 
\half\omega^M \; \frac{ \zeta^M_{1} \zeta^M_{2} }{ \abs{\bm{\xi}^M}^3 } \Delta V_M \\ \\

\half\omega^1 \; \frac{ \zeta^1_{1} \zeta^1_{3} }{ \abs{\bm{\xi}^1}^3 } \Delta V_1 &  
\half\omega^2 \; \frac{ \zeta^2_{1} \zeta^2_{3} }{ \abs{\bm{\xi}^2}^3 } \Delta V_2	& 
\half\omega^3 \; \frac{ \zeta^3_{1} \zeta^3_{3} }{ \abs{\bm{\xi}^3}^3 } \Delta V_3 & \cdots & 
\half\omega^M \; \frac{ \zeta^M_{1} \zeta^M_{3} }{ \abs{\bm{\xi}^M}^3 } \Delta V_M \\ 

\vdots & \vdots & \vdots & \vdots & \vdots\\

\half\omega^1 \; \frac{ \zeta^1_{6} \zeta^1_{6} }{ \abs{\bm{\xi}^1}^3 } \Delta V_1 &  
\half\omega^2 \; \frac{ \zeta^2_{6} \zeta^2_{6} }{ \abs{\bm{\xi}^2}^3 }\Delta V_2	& 
\half\omega^3 \; \frac{ \zeta^3_{6} \zeta^3_{6} }{ \abs{\bm{\xi}^3}^3 } \Delta V_3 & \cdots & 
\half\omega^M \; \frac{ \zeta^M_{6} \zeta^M_{6} }{ \abs{\bm{\xi}^M}^3 } \Delta V_M
\end{bmatrix}
\begin{bmatrix} 
c^1\\ \\
c^2\\ \\
c^3\\ 
\vdots\\
c^M 
\end{bmatrix}, 
\end{equation}

\noindent or in matrix notation can be written as, 
\begin{equation}
\mathcal{\bf{\tilde{C}}}^{PD} = \mathcal{\bf{X}}\bm{c}, \label{disc_stiff_tens}
\end{equation}

\noindent where tilde is used to denote that the stiffness components are written in vector form (containing terms in the voigt stiffness matrix). 

If the desired elastic stiffness of the material or the `reference' stiffness is similarly denoted by $\mathcal{\bf{\tilde{C}}}^{ref}$, then ideally, the aim would be to find a solution $\bm{c}$ such that $ \mathcal{\bf{\tilde{C}}}^{PD} = \mathcal{\bf{\tilde{C}}}^{ref}$. In general, for all practical purposes this coefficient matrix is rectangular and wide. Therefore we seek to find a solution $\bm{c}$ that has the minimum error in a least-squares sense, that is,
\begin{equation}
\bm{c} = \bm{c}: c_i \geq 0,  \small{min} \; \norm {\mathcal{\bf{\tilde{C}}}^{ref} - \mathcal{\bf{X}}\bm{c}}^2. \label{ls_1}
\end{equation} 

\noindent However, the least-squares solution does not guarantee non-negativity of all components of the solution $\bm{c}$. This requirement of non-negativity of the micromoduli stems from the fact that the stored strain energy density of the bond given in Eq. \eqref{Energy_5} has to be non-negative, $w(\bm{\xi,\eta}) \geq 0$ for any deformation which implies $c \geq 0$. Therefore, an additional constraint needs to be imposed - all components of the solution vector $c$ have to be non-negative. 

One way to find a unique solution to the problem is find that solution to the least-squares problem which also gives the lowest possible values of micromoduli. In other words, to find the minimum norm least-squares solution where in addition to minimizing the residual given by $\norm {\mathcal{\bf{\tilde{C}}}^{ref} - \mathcal{\bf{X}}\bm{c}}^2$, the norm of the solution $\norm{\bm{c}}$ itself is minimized. This can be computed from the Moore-Penrose pseudoinverse of $\bf{X}$ (in a software such as MATLAB\textsuperscript{\textregistered}, the functions \textit{pinv} or \textit{lsqminnorm} can be used to compute this solution). However, once again, this does not guarantee non-negativity of the solution $c$. Therefore to find the solution with the least norm and least residual within the space of positive solutions, the following scheme is proposed. 

Let $\bm{v}$ be a vector of size M $\times$ 1 such that it lies in the null space  of $\mathcal{\bf{X}}$ such that,
\begin{equation}
\mathcal{\bf{X}}\bm{v} = \bm{0}.
\end{equation}

\noindent where, $\bf{X}$ is the coefficient matrix in Eq. \eqref{disc_stiff_tens}. Then the desired solution is the vector $\bm{c + v}$ such that the norm $\norm{\bm{c + v}}$ is the minimum among all possible non-negative solutions,  $\bm{c + v \geq 0}$. In other words, the desired solution is obtained by a two-step process:

\begin{enumerate}
	\item First, evaluate the minimum norm least-squares solution $\bm{c}$.
	\item Second, this solution $\bm{c}$ is perturbed by an amount $\bm{v}$ such that $\bm{c + v}$ is non-negative and such that the residual $\norm {\mathcal{\bf{\tilde{C}}}^{ref} - \mathcal{\bf{X}}\bm{(c + v)}}^2$ remains unchanged since $\bm{v}$ lies in the null space of $\mathcal{\bf{X}}$.
\end{enumerate}

This scheme can be posed as a quadratic program subject to linear equality and inequality constraints as follows. Finding $min\, \norm{\bm{c + v}}$ is equivalent to finding the $min\, \half\bm{(c + v)}^T\bm{(c + v)}$ which can be expanded as, 
\begin{align}
min\, \half\bm{(c + v)}^T \bm{(c + v)} &= min\, \half\left( \bm{c^Tc} + \bm{v^Tc} + \bm{c^Tv} + \bm{v^Tv}\right), \label{quad_1}\\
&= min\, \half \bm{v^Tv} + \bm{c^Tv},\label{quad_2}\\
&= min\, \half\bm{v^TIv} + \bm{c^Tv},\label{quad_3}
\end{align}

\noindent where $\bm{I}$ is an M $\times$ M identity matrix and the term $\bm{c^Tc}$ is dropped since it is a known positive constant and does not affect the minimum. Therefore, the quadratic program can be now written as, 
\begin{align}
Minimize \;\; &\half\bm{v^TIv} + \bm{c^Tv}; \label{min_1} \\
Subject\; to \;\; &\bm{c} + \bm{v} \geq \bm{0}, \label{min_2} \\
& \mathcal{\bf{X}}\bm{v} = \bm{0}. \label{min_3}	 
\end{align}

Note that in the first step, other routines such as \textit{lsqlin} and \textit{lsqnonneg} can also be used to obtain the original solution $\bm{c}$, which may return one of multiple solutions for $\bm{c}$. For example, using \textit{lsqnonneg} returns a solution of non-negative values in the solution vector but not necessarily of minimum norm. However, this is of no consequence as the second step aims to find the minimum norm solution $\bm{(c + v)}$ which is cast as the objective function to be minimized in the quadratic program. The minor advantage of using say, \textit{lsqminnorm} in the first step is that if for a particular case, this routine finds the minimum norm solution with all non-negative micromoduli, then that is \textit{the} desired solution and the need for the second step is eliminated.

Although a formal proof of existence of solutions is not attempted here, as will be seen in the forthcoming results, solutions are found for a very wide variety of practical cases thus demonstrating the generality and robustness of the proposed method.

\section{Results and Discussion}
Results are given here to demonstrate the feasibility of this approach in producing a distribution of bond micromoduli for anisotropic bond-based models. Anisotropy is generally defined by a symmetry group for a material, a group of orthogonal transformations under which the fourth order elasticity tensor for a material is invariant. Simply put, suppose that a material is defined by a fourth order elasticity tensor $C_{ijkl}$, such that if under an orthogonal transformation defined by a matrix $Q_{ij}$,
\begin{equation}
C'_{pqrs} = Q_{pi}Q_{qj}Q_{rk}Q_{sl}C_{ijkl}, \label{sym_def}
\end{equation}

\noindent if $C'_{pqrs} = C_{ijkl}$, then $Q_{ij}$ is said to be a symmetry transformation of $C_{ijkl}$. In addition, a material is said to possess a \textit{symmetry plane} if there exists a \textit{reflection} transformation that satisfies Eq. \eqref{sym_def} such that $Q_{ij}$ can be defined as \cite{ting1996anisotropic},
\begin{equation}
Q_{ij} = \bm{R}(\bm{n}) = I_{ij} - 2n_i n_j \label{refl}
\end{equation}

\noindent where $\bm{n}$ is the unit normal to the plane of reflection. It turns out that all eight material symmetries can be obtained by successive introduction of symmetry planes starting from the most anisotropic - triclinic to the least - isotropic \cite{chadwick2001new, cowin1987identification}. The group $\mathcal{S}$ of all possible orthogonal transformations $Q_{ij}$ for a material under which the material's elasticity tensor remains invariant is known as the symmetry group for that material. 

Note that all anisotropic materials admit the identity transformation $\bm{I}$ and the central inversion transformation $-\bm{I}$ \cite{ting1996anisotropic} where,
\begin{equation}
\bm{I}=
\begin{bmatrix} 
1 & 0 & 0 \\
0 & 1 & 0 \\
0 & 0 & 1 \\
\end{bmatrix},
\bm{-I}=
\begin{bmatrix} 
-1 & 0 & 0 \\
0 & -1 & 0 \\
0 & 0 & -1 \\
\end{bmatrix}.
\end{equation}

The peridynamic stiffness tensor Eq. \eqref{Cijkl_5} of course admits these transformations as well. Notably, the inversion transformation can be written from Eq. \eqref{Cijkl_5} and Eq. \eqref{sym_def} as,

\begin{equation}
\mathcal{\bf{C}}^{PD} = \half \int_{H_{\bm{x}}} \ifun c \; \frac{\bm{\xi} \otimes \bm{\xi} \otimes \bm{\xi} \otimes \bm{\xi}}{ \blen^3 } \dvol = \half \int_{H_{\bm{x}}} \ifun c \; \frac{\bm{-\xi} \otimes \bm{-\xi} \otimes \bm{-\xi} \otimes \bm{-\xi}}{ \blen^3 } \dvol. \label{Cijkl_inv} 
\end{equation} 

\noindent In other words, the contribution of individual bonds to the peridynamics stiffness tensor is invariant if the bond is reflected about it's own axis and therefore the micromoduli $c\left(\bm{\xi}\right)=c\left(-\bm{\xi}\right)$. This can be observed for all cases in the forthcoming results.

All eight possible symmetries for linear elastic materials are tested with material properties taken from \cite{trageser2019anisotropic} (except for triclinic which is obtained from \cite{doi:10.1038/sdata.2015.9}). Not all of these materials satisfy Cauchy's relations \textit{exactly} (but nearly), however are still used as inputs as is. This is done deliberately so that first, real materials may be used as examples, second so that errors in the calibration may be estimated when using realistic material properties which may not satisfy exactly Cauchy's relations and finally so that the reader may observe that Cauchy's relations emerge naturally in the calibrated elastic stiffness tensor.

It is important to note that while any material stiffness tensor can be used to derive bond micromoduli distributions using the proposed method, it is most effective when using materials that exactly satisfy Cauchy's relations. This is of course because the bond-based peridynamics model itself, due to the assumption of pairwise interactions, generates an elasticity tensor which inherently satisfies Cauchy's relations.

The computational least-squares method is implemented in the software, MATLAB\textsuperscript{\textregistered}. The discretized peridynamic neighborhood or `lattice' is assumed to be a 3D regular grid with each particle imagined to be a cube of unit size, unit volume, and therefore has unit lattice spacing ($\Delta = 1$). The lattice is assumed to be in the bulk, i.e. away from any surface or edges. The lattice and the material's coordinate system are assumed to be oriented with each other in the same Cartesian coordinate system with basis $\{\bm{e_1}, \bm{e_2}, \bm{e_3}\}$ \footnote{Coordinates are also referred to as {x,y,z} in the text for ease and brevity} unless otherwise stated. A spherical neighborhood (meaning particle falling within a sphere, similar a von Neumann neighborhood in 3D) with a horizon radius of 6 times the lattice spacing (i.e $\delta = 6\Delta = 6$), and an inverse influence function, i.e. $\ifun = \frac{\delta}{\abs{\bm{\xi}}}$ is used for all examples unless otherwise stated. All elastic constants have units of GPa. For the sake of completeness, the degree of anisotropy for each case is also evaluated and presented in the form of the universal anisotropy index \cite{ranganathan2008universal, doi:10.1038/sdata.2015.9} (see \ref{UAI}). 

\subsection{Eight symmetries of linear elastic materials} \label{section_8_symm}
It is well known that there are eight possible symmetries for linear elastic material - triclinic, monoclinic, orthotropic, trigonal, tetragonal, transversely isotropic, cubic and isotropic \cite{ting1996anisotropic, chadwick2001new, trageser2019anisotropic}. Presented below is an example of each case starting with triclinic to isotropic in relatively decreasing order of anisotropy.  

\subsubsection{Triclinic Symmetry}
Triclinic, the most general material class has no planes of symmetry, i.e. all 21 constants in the upper triangle of the Voigt stiffness matrix can be non-zero and are independent. The symmetry group for triclinic materials can be written quite simply as,
\begin{equation}
\mathcal{S}=\{ \bm{I}, \bm{-I}\}.
\end{equation}

With Cauchy's restrictions imposed, the number of independent constants for a triclinic material reduces from 21 to 15, and can be visualized as follows,
\begin{equation}
\begin{bmatrix} 
C_{11}&    C_{12}&    C_{13}&	C_{14}&	C_{15}&	C_{16}\\
C_{21}&    C_{22}&    C_{23}&	C_{24}&	C_{25}&	C_{26}\\
C_{31}&    C_{32}&    C_{33}&	C_{34}&	C_{35}&	C_{36}\\
C_{41}&    C_{42}&    C_{43}&	C_{44}&	C_{45}&	C_{46}\\
C_{51}&    C_{52}&    C_{53}&	C_{54}&	C_{55}&	C_{56}\\
C_{61}&    C_{62}&    C_{63}&	C_{64}&	C_{65}&	C_{66}\\
\end{bmatrix}
\rightarrow
\begin{bmatrix} 
C_{11}&    	\odot&  \Diamond&	\otimes&	C_{15}&	C_{16}\\
\odot&    C_{22}&    	\Box&	C_{24}&	\ominus&	C_{26}\\
\Diamond&    \Box&    C_{33}&	C_{34}&	C_{35}&	\star\\
\otimes&    C_{42}&    C_{43}&	\Box&	\star&	\ominus\\
C_{51}&    \ominus&    C_{53}&	\star&	\Diamond&	\otimes\\
C_{61}&    C_{62}&    \star&	\ominus&	\otimes&	 \odot\\
\end{bmatrix}. \label{cauchy_mat_tric}
\end{equation}

Entries containing the same symbol are equal to each other. The other entries are left as is to differentiate them from the ones affected by Cauchy's relations. Note that for a triclinic material, 9 elastic constants remain unaffected while 6 out of the other 12 are left truly independent after Cauchy's relations are imposed, leading to a total of 15 independent elastic constants. An example of a triclinic material, KIO$_{3}$, is given below with properties are taken from \cite{doi:10.1038/sdata.2015.9}. The Voigt elastic stiffness matrix is given by,
\begin{equation}
\mathcal{\bf{\tilde{C}}}^{ref} = 
\begin{bmatrix} 
43&    11&    13&        1&         -2&         2 \\
11&    35&    12&       3&         -1&         3\\
13&    12&    43&      2&         -2&         1\\
1&     3&      2&    	13&      0&         0 \\
-2&    -1&    -2&    	0&        13&         1\\
2&      3&     1&   0&        1&    	12\\
\end{bmatrix}. \label{voigt_tric}
\end{equation}
\noindent It appears that some terms a priori satisfy Cauchy's relations exactly, for example $C_{13} = C_{55}$ and $C_{14} = C_{56}$ where as others are quite close resulting in an anisotropy coefficient $A^U$ of 0.1965.

\begin{figure}[H]
	\centering
	\captionsetup{font=footnotesize}
	\captionsetup{justification=centering}	
	\begin{subfigure}[t]{0.32\textwidth}
		\centering
		\includegraphics[width=1\textwidth]{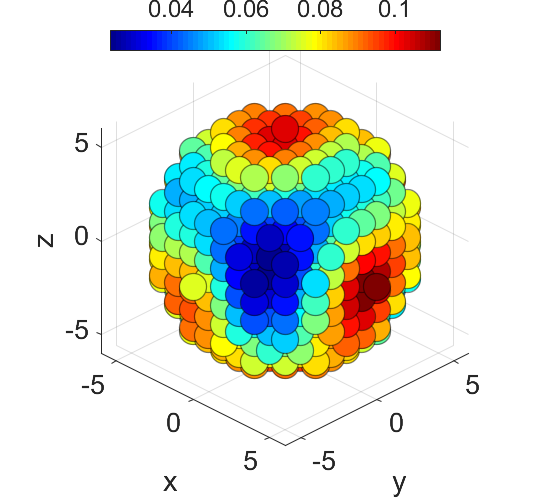}
		\caption{}
		\label{tric_1}
	\end{subfigure}%
	\begin{subfigure}[t]{0.32\textwidth}
		\centering
		\includegraphics[width=1\textwidth]{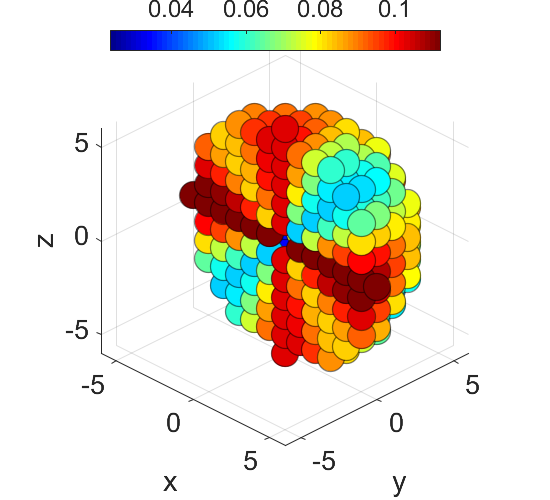}
		\caption{}
		\label{tric_2}
	\end{subfigure}
	\begin{subfigure}[t]{0.32\textwidth}
		\centering
		\includegraphics[width=1\textwidth]{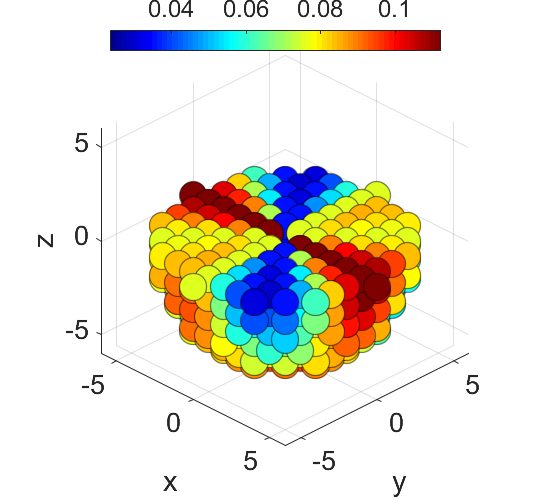}
		\caption{}
		\label{tric_3}
	\end{subfigure}
	\caption[]{(a) Complete neighborhood, and partial views of the neighborhood with normal to the cut plane at (b) $\bm{n} = -\bm{e_2}$ and (c)  $\bm{n} = \bm{e_3}$ for triclinic symmetry  with horizon $\delta = 6$ and influence function \( \ifun = \frac{\delta}{\abs{\bm{\xi}}} \).}
	\label{tric}
\end{figure}

The bond micromoduli are solved for using the least-squares method described in section \ref{opti}. For the purposes of illustration, the particles are depicted as colored spheres in Figure \ref{tric} with the color depicting the magnitude of bond micromoduli between a particle within the horizon in the lattice and the reference particle at the center of the lattice. Figure \ref{tric_1} shows the complete neighborhood, Figures \ref{tric_2} and \ref{tric_3} shows partial views of the neighborhood with normal to the cut plane at $\bm{n} = -\bm{e_2}$ and $\bm{n} = \bm{e_3}$ respectively so that the reader may observe the distribution of bonds inside the volume of the neighborhood better. True to the triclinic material class, there appear to be no symmetries present in the peridynamic neighborhood from a visual inspection of Figure \ref{tric} (other than the inversion symmetry). Bonds oriented in the $x$ and $z$ direction appear to be of higher stiffness than those oriented in the $y$ direction in general, which appears to be qualitatively correct from looking at Eq. \eqref{voigt_tric}.

The calibrated peridynamic stiffness tensor which can be obtained by $\mathcal{\bf{\tilde{C}}}^{PD} = \mathcal{\bf{X}}\left(\bm{c+v}\right)$ is presented in Voigt notation in Eq. \eqref{voigt_tric_pd}.
\begin{equation}
\mathcal{\bf{\tilde{C}}}^{PD} = 
\begin{bmatrix} 
43&    11.67&    13&        1&         -2&         2 \\
11.67&    35&    12.67&       3&         -0.33&         3\\
13&    12.67&    43&      2&         -2&         0.33\\
1&     3&      2&    	12.67&      0.33&         -0.33\\
-2&    -0.33&    -2&    	0.33&        13&         1\\
2&      3&     0.33&   -0.33&        1&    	11.67\\
\end{bmatrix}. \label{voigt_tric_pd}
\end{equation}

\noindent The 9 entries which remain unaffected by Cauchy's relations, namely $C_{11}$, $C_{22}$, $C_{33}$, $C_{24}$, $C_{34}$, $C_{35}$, $C_{15}$, $C_{16}$, $C_{26}$ are left unchanged during the least-squares calibration and the other 12 are modified (compare the structure of the matrices given in Eq. \eqref{voigt_tric_pd} and Eq. \eqref{cauchy_mat_tric}). 

Note that the reference stiffness matrix in Eq. \eqref{voigt_tric} is used as input to the model as is, and not modified in any way to satisfy Cauchy's relations a priori. Cauchy's relations are found to emerge naturally as part of the least-squares calibration procedure. The relative error in the solution, which is defined here as $\norm{ \mathbb{\bf{\tilde{C}}}^{ref} - \mathbb{\bf{\tilde{C}}}^{PD}}/\norm{\mathbb{\bf{\tilde{C}}}^{ref}}$ is 3.1873\% where $\norm{\cdot}$ represents the $L_2$ norm.  

\subsubsection{Monoclinic Symmetry}\label{section_mono}
Monoclinic symmetry has a single plane of symmetry, which in the present case is the $x-y$ plane or $z=0$ plane with the symmetry group (borrowing notation from Eq. \eqref{refl}),
\begin{equation}
\mathcal{S}=\{ \bm{I}, \bm{-I}, \bm{R}(\bm{e_3}) \}.
\end{equation}

\noindent This gives rise to 13 independent constants namely,
\begin{equation}
C_{11}, C_{22}, C_{33}, C_{44}, C_{55}, C_{66}, C_{12}, C_{13}, C_{23}, C_{16}, C_{26}, C_{36}, C_{45}. 
\end{equation}

\noindent Cauchy's relations reduce the number of independent constants from 13 to 9,
\begin{equation}
C_{11}, C_{22}, C_{33}, C_{44} = C_{23}, C_{55} = C_{13}, C_{66} = C_{12}, C_{16}, C_{26}, C_{36} = C_{45}, 
\end{equation}

\noindent and can be visualized as follows,
\begin{equation}
\begin{bmatrix} 
C_{11}&    C_{12}&    C_{13}&	0&	0&	C_{16}\\
C_{21}&    C_{22}&    C_{23}&	0&	0&	C_{26}\\
C_{31}&    C_{32}&    C_{33}&	0&	0&	C_{36}\\
0&    0&   0&	C_{44}&	C_{45}&	0\\
0&    0&   0&	C_{54}&	C_{55}&	0\\
C_{61}&    C_{62}&    C_{63}&	0&	0&	C_{66}\\
\end{bmatrix}
\rightarrow
\begin{bmatrix} 
C_{11}&    	\odot&  \Diamond&	0&	0&	C_{16}\\
\odot&    C_{22}&    	\Box&	0&	0&	C_{26}\\
\Diamond&    \Box&    C_{33}&	0&	0&	\star\\
0&    0&    0&	\Box&	\star&	0\\
0&    0&    0&	\star&	\Diamond&	0\\
C_{61}&    C_{62}&    \star&	0&	0&	 \odot\\
\end{bmatrix}. \label{cauchy_mat_mono}
\end{equation}

The example material chosen is CoTeO$_{4}$, with a universal anisotropy index of 28.3169 and Voigt elastic stiffness matrix given by,
\begin{equation}
\mathcal{\bf{\tilde{C}}}^{ref} = 
\begin{bmatrix} 
135&    19&    54&        0&         0&         42 \\
19&    13&    15&       0&         0&         6 \\
54&    	 15&    269&      0&         0&         18 \\
0&         0&       0&    	14&      25&         0 \\
0&         0&       0&    	25&        66&         0 \\
42&        6&       18&   0&        0&    	18\\
\end{bmatrix}. \label{voigt_mono}
\end{equation}

\begin{figure}[H]
	\centering
	\captionsetup{font=footnotesize}
	\captionsetup{justification=centering}	
	\begin{subfigure}[t]{0.32\textwidth}
		\centering
		\includegraphics[width=1\textwidth]{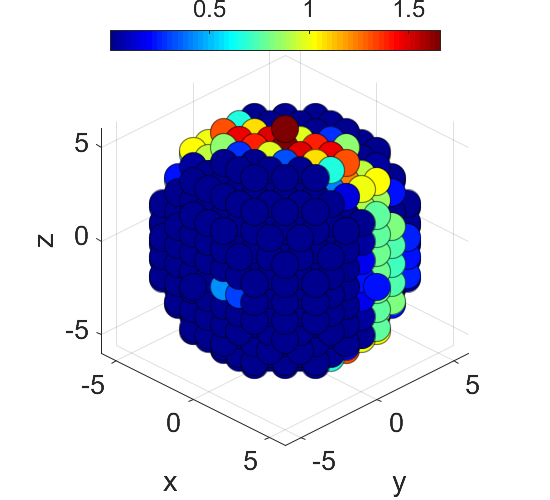}
		\caption{}
		\label{mono_1}
	\end{subfigure}%
	\begin{subfigure}[t]{0.32\textwidth}
		\centering
		\includegraphics[width=1\textwidth]{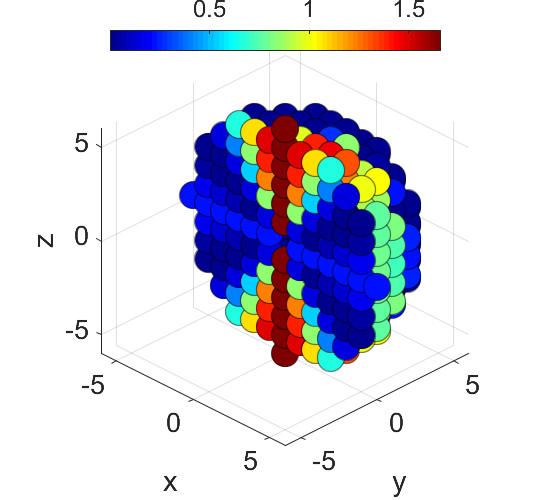}
		\caption{}
		\label{mono_2}
	\end{subfigure}
	\begin{subfigure}[t]{0.32\textwidth}
		\centering
		\includegraphics[width=1\textwidth]{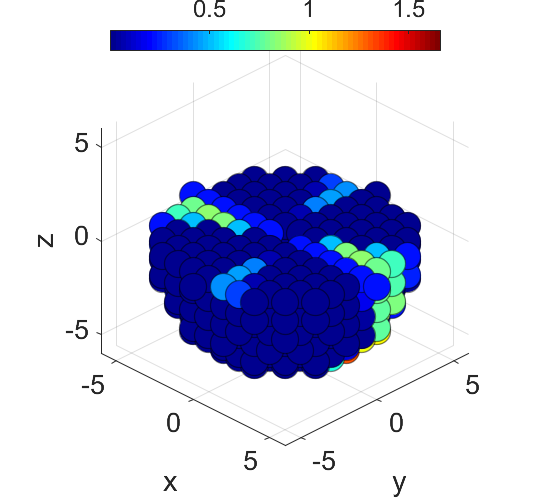}
		\caption{}
		\label{mono_3}
	\end{subfigure}
	\caption[]{(a) Complete neighborhood, and partial views of the neighborhood with normal to the cut plane at (b) $\bm{n} = -\bm{e_2}$ and (c)  $\bm{n} = \bm{e_3}$ for monoclinic symmetry  with horizon $\delta = 6$ and influence function \( \ifun = \frac{\delta}{\abs{\bm{\xi}}} \).}
	\label{mono}
\end{figure}

From Figure \ref{mono}, it is seen that the neighborhood contains only one plane of symmetry - the $x-y$ plane with the distribution of micromoduli reflected exactly across this plane. There are several bonds with micromoduli equal to zero, which points to the wide range of values in the material stiffness tensor, e.g. $C_{22} = 13 GPa$ vs. $C_{33} = 269 GPa$. The wide range of axial stiffness terms in orthogonal directions results in a lumped distribution of micromoduli. Bonds with the highest micromoduli are aligned preferentially with the $z$-axis followed by the $x$-axis, not surprising as $C_{33}$ and $C_{11}$ have the highest values in the material stiffness tensor. Micromoduli for many bonds not oriented perfectly in the $x-z$ plane are zero and most of the stiffness in the $y$ direction is carried by a few bonds along the $y$-axis. While this is not ideal in a realistic modeling scenario, one workaround could be to set a lower bound on the micromoduli in the second step of the calibration procedure, i.e. specify $\bm{c} + \bm{v} \geq \bm{l}$ in Eq. \eqref{min_2} where $\bm{l}$ is a vector of the minimum value of micromoduli desired. However, this vector $\bm{l}$ will have to chosen arbitrarily by trial and error since any desired value might not produce a solution that satisfies all the constraints in the calibration procedure.
\begin{equation}
\mathcal{\bf{\tilde{C}}}^{PD} = 
\begin{bmatrix} 
135&    18.33&    62&        0&         0&         42 \\
18.33&    13&    14.33&       0&         0&         6 \\
62&    	 14.33&    269&      0&         0&         22.67 \\
0&         0&       0&    	14.33&      22.67&         0 \\
0&         0&       0&    	22.67&        62&         0 \\
42&        6&       22.67&   0&        0&    	18.33\\
\end{bmatrix}. \label{voigt_mono_pd}
\end{equation}

The corresponding peridynamic stiffness tensor given by Eq. \eqref{voigt_mono_pd} is calibrated to within 4.988\% relative error. Note that once again that the 5 constants $C_{11}$, $C_{22}$, $C_{33}$, $C_{16}$ and $C_{26}$ that are not affected by Cauchy's relations remain unchanged while the others (denoted by symbols in Eq. \eqref{cauchy_mat_mono}) which are modified.

\subsubsection{Orthotropic Symmetry} \label{section_ortho}
Orthotropic symmetry can be generated by specifying two additional symmetry planes such that the symmetry group is given by,
\begin{equation}
\mathcal{S}=\{ \bm{I}, \bm{-I}, \bm{R}(\bm{e_1}), \bm{R}(\bm{e_2}), \bm{R}(\bm{e_3}) \}.
\end{equation}

Orthotropic symmetry is commonly encountered when modeling fiber reinforced composite materials, perhaps the class of anisotropy that is most commonly found in peridynamics literature. Orthotropic symmetry specifies that the material stiffness tensor can have 9 independent constants given by,
\begin{equation}
C_{11}, C_{22}, C_{33}, C_{44}, C_{55}, C_{66}, C_{12}, C_{13}, C_{23},
\end{equation} 

\noindent with all others being zero. Cauchy's relations reduce the number of independent constants from 9 to 6 (c.f. Azdoud et al. \cite{azdoud2013morphing}),
\begin{equation}
C_{11}, C_{22}, C_{33}, C_{44} = C_{23}, C_{55} = C_{13}, C_{66} = C_{12},
\end{equation} 

\noindent which can be visualized as follows,
\begin{equation}
\begin{bmatrix} 
C_{11}&    C_{12}&    C_{13}&	0&	0&	0\\
C_{21}&    C_{22}&    C_{23}&	0&	0&	0\\
C_{31}&    C_{32}&    C_{33}&	0&	0&	0\\
0&    0&   0&	C_{44}&	0&	0\\
0&    0&   0&	0&	C_{55}&	0\\
0&    0&    0&	0&	0&	C_{66}\\
\end{bmatrix}
\rightarrow
\begin{bmatrix} 
C_{11}&    	\odot&  \Diamond&	0&	0&	0\\
\odot&    C_{22}&    	\Box&	0&	0&	0\\
\Diamond&    \Box&    C_{33}&	0&	0&	0\\
0&    0&    0&	\Box&	0&	0\\
0&    0&    0&	0&	\Diamond&	0\\
0&    0&    0&	0&	0&	 \odot\\
\end{bmatrix}. \label{cauchy_mat_ortho}
\end{equation}

\noindent The example Voigt elastic stiffness matrix, for Te$_2$W, is given by,
\begin{equation}
\mathcal{\bf{\tilde{C}}}^{ref} = 
\begin{bmatrix} 
143&    1&    37&       0&         0&         	0 \\
1&    	3&    3&        0&         0&         		0 \\
37&    	3&    102&      0&         0&         	0 \\
0&         0&         	0&    	2&      0&         0 \\
0&         0&         	0&    	0&      46&         0 \\
0&         0&       	0&      0&        0&    1\\
\end{bmatrix}. \label{voigt_ortho}
\end{equation}

The distribution of calibrated bond micromoduli is given in Figure \ref{ortho} with the corresponding peridynamic stiffness tensor in Eq. \eqref{voigt_ortho_pd}. Figure \ref{ortho_2} quite clearly shows the orthotropic nature of the neighborhood although it is limited to a plane near the center. The bonds preferentially aligned with the $x$--axis have the highest micromoduli followed by the $z$-axis. Similar to the monoclinic material considered, the elastic constants span a wide range of values, e.g. from 3 in $C_{22}$ to 143 in $C_{11}$ giving a high universal anisotropy constant of 54.0623.

\begin{figure}[H]
	\centering
	\captionsetup{font=footnotesize}
	\captionsetup{justification=centering}	
	\begin{subfigure}[t]{0.32\textwidth}
		\centering
		\includegraphics[width=1\textwidth]{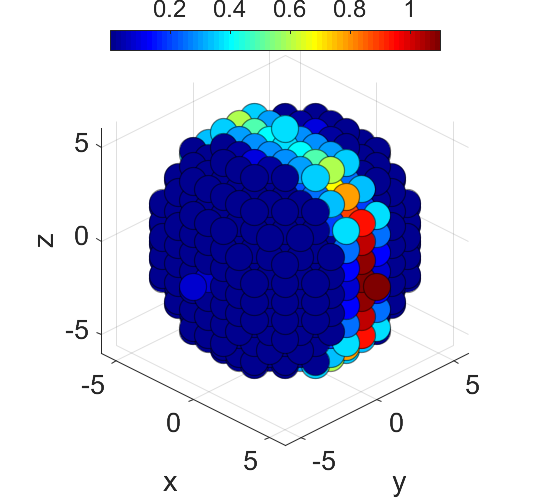}
		\caption{}
		\label{ortho_1}
	\end{subfigure}%
	\begin{subfigure}[t]{0.32\textwidth}
		\centering
		\includegraphics[width=1\textwidth]{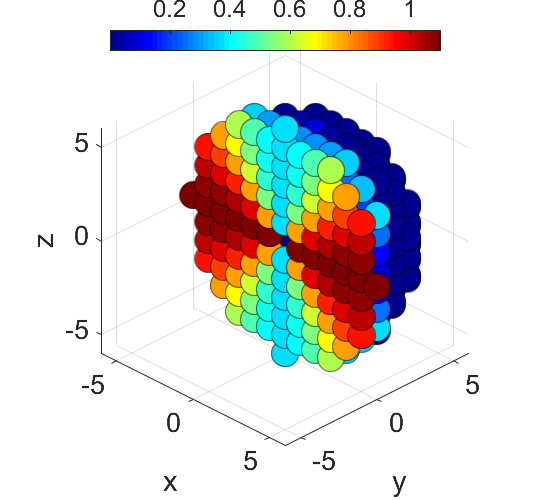}
		\caption{}
		\label{ortho_2}
	\end{subfigure}
	\begin{subfigure}[t]{0.32\textwidth}
		\centering
		\includegraphics[width=1\textwidth]{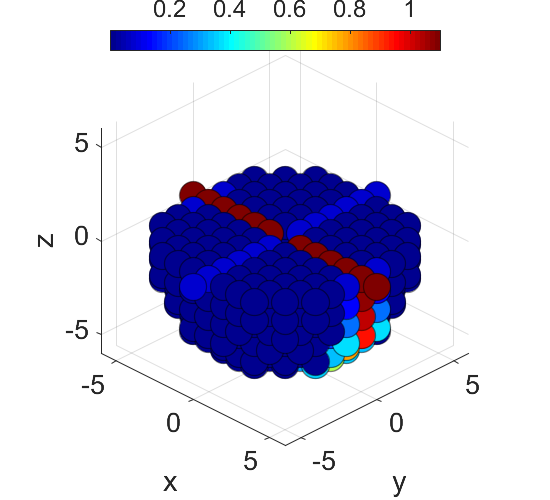}
		\caption{}
		\label{ortho_3}
	\end{subfigure}
	\caption[]{(a) Complete neighborhood, and partial views of the neighborhood with normal to the cut plane at (b) $\bm{n} = -\bm{e_2}$ and (c)  $\bm{n} = \bm{e_3}$ for orthotropic symmetry  with horizon $\delta = 6$ and influence function \( \ifun = \frac{\delta}{\abs{\bm{\xi}}} \).}
	\label{ortho}
\end{figure}

The calibrated peridynamics stiffness tensor is given in Eq. \eqref{voigt_ortho_pd} for which the relative error is found to be 5.0958\%. Note that $C_{12} = C_{66}$ a priori satisfy Cauchy's relations exactly which help reduce the error in this case. As with previous cases, the axial stiffness terms remain unchanged where as the 6 other terms namely the off-diagonal axial terms as well as the shear stiffnesses are modified during the calibration. 
\begin{equation}
\mathcal{\bf{\tilde{C}}}^{PD} = 
\begin{bmatrix} 
143&    1&    43&       0&         0&         	0 \\
1&    	3&    2.33&        0&         0&         		0 \\
43&    	2.33&    102&      0&         0&         	0 \\
0&         0&         	0&    	2.33&      0&         0 \\
0&         0&         	0&    	0&      43&         0 \\
0&         0&       	0&      0&        0&    1\\
\end{bmatrix}. \label{voigt_ortho_pd}
\end{equation}

\subsubsection{Trigonal Symmetry}
Trigonal symmetry may be generated by choosing three planes of symmetry as well, similar to orthotropic symmetry, with the normals to the planes lying in the $x-z$ plane such that the symmetry group can be written as,
\begin{equation}
\mathcal{S}=\Bigg\{ \bm{I}, \bm{-I}, \bm{R}\left(\bm{e_1}\right), \bm{R}\left(\frac{\sqrt{3}}{2} \bm{e_1} \pm \frac{1}{2} \bm{e_2}\right) \Bigg\}.
\end{equation}

\noindent The restrictions on the non-zero elastic constants are,
\begin{equation}
C_{11} = C_{22}, \; C_{13} = C_{23}, \; C_{33}, \; C_{12}, \; C_{44} = C_{55}, \; C_{14} = C_{56} = - C_{14}, \; C_{66} = (C_{11} - C_{12})/2. \label{tri_conditions}
\end{equation}

Note that along with other restrictions, Cauchy's relations notably impose $C_{66}=C_{12}$ which combined with the last of Eq. \eqref{tri_conditions} imposes,

\begin{equation}
(C_{11} - C_{12})/2 = C_{12} \implies C_{11}=3C_{12}. \label{tri_66_12}
\end{equation}

Interestingly Cauchy's symmetry impose $C_{14} = C_{56}$ which trigonal symmetry by itself possesses. Therefore, combining Eq. \eqref{tri_conditions} and Eq. \eqref{tri_66_12} the number of independent constants reduces from 6 to 4, and can be written as follows,
\begin{equation}
C_{11} = C_{22} = 3C_{12} = 3C_{66}, \; C_{33}, \; C_{13} = C_{23} = C_{44} = C_{55}, \; C_{14} = C_{56} = - C_{24}, \label{tri_conditions_cauchy}
\end{equation}

\noindent and can be visualized as follows\footnote{It is implied that the matrices given in Eq. \eqref{cauchy_mat_mono} are symmetric, therefore $C_{JI} = C_{IJ}$ and the elastic constants $C_{IJ}$ that are not denoted by symbols satisfy the constraints of the particular material class chosen. For example, it is implied in Eq. \eqref{tri_conditions_cauchy} that $C_{22}=C_{11}$ and they are not truly independent of each other.},
\begin{equation}
\begin{bmatrix} 
C_{11}&    C_{12}&    C_{13}&	C_{14}&	0&	0\\
C_{21}&    C_{22}&    C_{23}&	C_{24}&	0&	0\\
C_{31}&    C_{32}&    C_{33}&	0&	0&	0\\
C_{41}&    C_{42}&   0&	C_{44}&	0&	0\\
0&    0&   0&	0&	C_{55}&	C_{56}\\
0&    0&    0&	0&	C_{65}&	C_{66}\\
\end{bmatrix}
\rightarrow
\begin{bmatrix} 
C_{11}&    	\odot&  \Diamond&	\otimes&	0&	0\\
\odot&    C_{22}&    	\Box&	\otimes&	0&	0\\
\Diamond&    \Box&    C_{33}&	0&	0&	0\\
\otimes&    \otimes&    0&	\Box&	0&	0\\
0&    0&    0&	0&	\Diamond&	\otimes\\
0&    0&    0&	0&	\otimes&	 \odot\\
\end{bmatrix}. \label{cauchy_mat_tri}
\end{equation}

\noindent Note that the $\otimes$ symbol in the $C_{24}$ position denotes that the entry is related but not equal (sign is opposite that of $C_{12}$ and $C_{56}$). The Voigt elastic stiffness matrix, with a universal anisotropy index of $A^U = 0.5875$ (for Ta$_2$C) is given by,
\begin{equation}
\mathcal{\bf{\tilde{C}}}^{ref} = 
\begin{bmatrix} 
464&    159&    141&       -45&         0&         	0 \\
159&    464&    141&        45&         0&         	0 \\
141&    141&    493&         0&         0&          0 \\
-45&     45&      0&    125&      0&         0 \\
0&         0&     0&      0&      125&         -45 \\
0&         0&     0&      0&       -45&    153\\
\end{bmatrix}. \label{voigt_tri}
\end{equation}

\begin{figure}[H]
	\centering
	\captionsetup{font=footnotesize}
	\captionsetup{justification=centering}	
	\begin{subfigure}[t]{0.32\textwidth}
		\centering
		\includegraphics[width=1\textwidth]{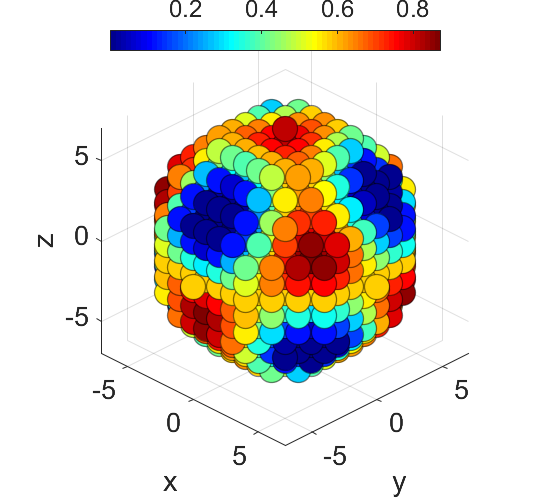}
		\caption{}
		\label{tri_1}
	\end{subfigure}%
	\begin{subfigure}[t]{0.32\textwidth}
		\centering
		\includegraphics[width=1\textwidth]{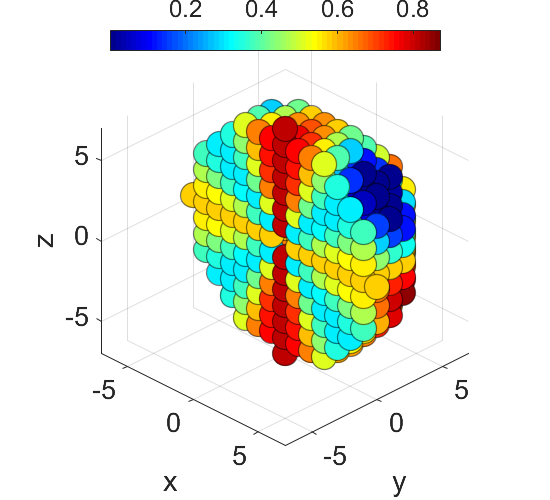}
		\caption{}
		\label{tri_2}
	\end{subfigure}
	\begin{subfigure}[t]{0.32\textwidth}
		\centering
		\includegraphics[width=1\textwidth]{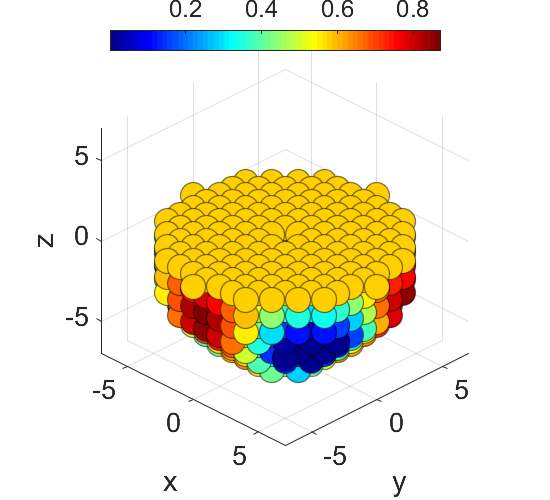}
		\caption{}
		\label{tri_3}
	\end{subfigure}
	\caption[]{(a) Complete neighborhood, and partial views of the neighborhood with normal to the cut plane at (b) $\bm{n} = -\bm{e_2}$ and (c)  $\bm{n} = \bm{e_3}$ for trigonal symmetry  with horizon $\delta = 6$ and influence function \( \ifun = \frac{\delta}{\abs{\bm{\xi}}} \).}
	\label{tri}
\end{figure}

The resulting distribution of bonds is given in Figure \ref{tri}. Note that from visual inspection of Figure \ref{tri} the peridynamic neighborhood is not symmetric about the $y-z$ plane or the $x-z$ plane but is still symmetric about the $x-y$ plane. The calibrated peridynamic stiffness tensor is given in Eq. \eqref{voigt_tri_pd}. As in previous cases, the axial diagonal terms are captured exactly in addition to the axial-shear coupling terms $C_{14}, C_{24}$ and shear coupling term $C_{56}$. The calibrated values of $C_{66}$ and $C_{12}$ of 155 are very close to their reference values as well, with the largest difference stemming from the axial-coupling terms $C_{13}, C_{23}$ and the shear terms $C_{44}, C_{55}$. Although there are 5 unique elastic constants appearing in effective peridynamic stiffness matrix Eq. \eqref{voigt_tri_pd}, $C_{11}=C_{22}$ are related to $C_{12}=C_{66}$ as noted before ($C_{66} \approx \frac{1}{3}C_{11}$). The overall relative error in the calibration is 2.632\%.
\begin{equation}
\mathcal{\bf{\tilde{C}}}^{PD} = 
\begin{bmatrix} 
464&    155&    130.33&       -45&         0&         	0 \\
155&    464&    130.33&        45&         0&         	0 \\
130.33& 130.33&    493&         0&         0&         	0 \\
-45&        45&         0&    130.33&       0&         0 \\
0&         0&         0&         0&    130.33&        -45 \\
0&         0&         0&         0&         -45&    155\\
\end{bmatrix}. \label{voigt_tri_pd}
\end{equation}

\subsubsection{Tetragonal Symmetry}

Tetragonal symmetry can be imagined to be orthotropic symmetry with two additional planes of reflection, with a symmetry group given by,
\begin{equation}
\mathcal{S}=\Bigg \{ \bm{I}, \bm{-I}, \bm{R}(\bm{e_1}), \bm{R}(\bm{e_2}), \bm{R}(\bm{e_3}), \bm{R}\left(\frac{1}{\sqrt{2}} \bm{e_1} \pm \frac{1}{\sqrt{2}} \bm{e_2}\right) \Bigg \}.
\end{equation}

\noindent Tetragonal symmetry specifies that the elastic stiffness tensor can have 6 independent constants given by,
\begin{equation}
C_{11} = C_{22}, C_{33}, C_{44} = C_{55}, C_{66}, C_{12}, C_{13} = C_{23},
\end{equation}

\noindent with all others being zero. Cauchy's relations reduce the number of independent constants from 6 to 4 given by,
\begin{equation}
C_{11} = C_{22}, C_{33}, C_{44} = C_{55} = C_{13} = C_{23}, C_{66} = C_{12},
\end{equation} 

\noindent and can be visualized as follows,
\begin{equation}
\begin{bmatrix} 
C_{11}&    C_{12}&    C_{13}&	0&	0&	0\\
C_{21}&    C_{22}&    C_{23}&	0&	0&	0\\
C_{31}&    C_{32}&    C_{33}&	0&	0&	0\\
0&    0&   0&	C_{44}&	0&	0\\
0&    0&   0&	0&	C_{55}&	0\\
0&    0&    0&	0&	0&	C_{66}\\
\end{bmatrix}
\rightarrow
\begin{bmatrix} 
C_{11}&    	\odot&  \Box&	0&	0&	0\\
\odot&    C_{22}&    	\Box&	0&	0&	0\\
\Box&    \Box&    C_{33}&	0&	0&	0\\
0&    0&    0&	\Box&	0&	0\\
0&    0&    0&	0&	\Box&	0\\
0&    0&    0&	0&	0&	 \odot\\
\end{bmatrix}. \label{cauchy_mat_tet}
\end{equation}

The example Voigt elastic stiffness matrix, for Si, is given by,
\begin{equation}
\mathcal{\bf{\tilde{C}}}^{ref} = 
\begin{bmatrix} 
212&    70&    58&         0&         0&         0 \\
70&    212&    58&         0&         0&         0 \\
58&    58&    179&         0&         0&         0 \\
0&         0&         0&    58&         0&         0 \\
0&         0&         0&         0&    58&         0 \\
0&         0&         0&         0&         0&    85\\
\end{bmatrix}. \label{voigt_tet}
\end{equation}

\noindent Interestingly, this material partially satisfies Cauchy's relations as seen by the equality of $C_{13}, C_{23}, C_{44}, C_{55}$ which will help reduce the error in the least-squares calibration process. The universal anisotropy index for this material is given by $A^U = 0.1231$.
\begin{figure}[H]
	\centering
	\captionsetup{font=footnotesize}
	\captionsetup{justification=centering}	
	\begin{subfigure}[t]{0.32\textwidth}
		\centering
		\includegraphics[width=1\textwidth]{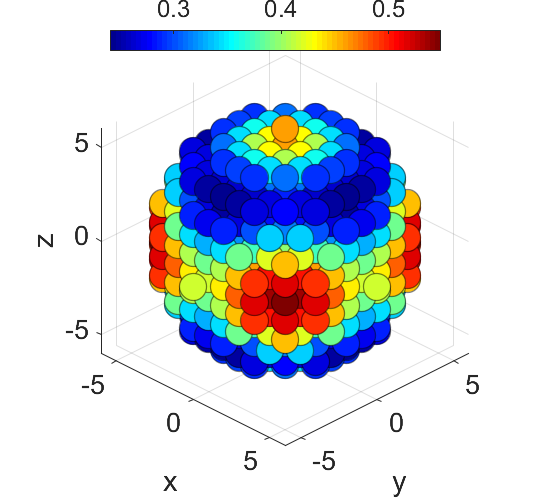}
		\caption{}
		\label{tet_1}
	\end{subfigure}%
	\begin{subfigure}[t]{0.32\textwidth}
		\centering
		\includegraphics[width=1\textwidth]{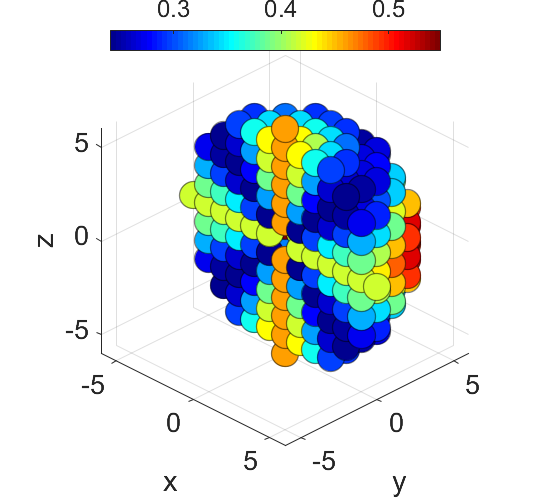}
		\caption{}
		\label{tet_2}
	\end{subfigure}
	\begin{subfigure}[t]{0.32\textwidth}
		\centering
		\includegraphics[width=1\textwidth]{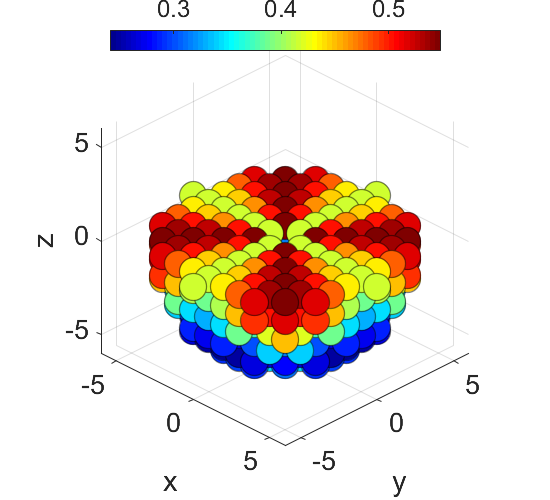}
		\caption{}
		\label{tet_3}
	\end{subfigure}
	\caption[]{(a) Complete neighborhood, and partial views of the neighborhood with normal to the cut plane at (b) $\bm{n} = -\bm{e_2}$ and (c)  $\bm{n} = \bm{e_3}$ for tetragonal symmetry  with horizon $\delta = 6$ and influence function \( \ifun = \frac{\delta}{\abs{\bm{\xi}}} \).}
	\label{tet}
\end{figure}

The solution is shown graphically in Figure \ref{tet} and the effective peridynamics stiffness tensor is given by, 
\begin{equation}
\mathcal{\bf{\tilde{C}}}^{PD} = 
\begin{bmatrix} 
212&    80&    58&         0&         0&         0 \\
80&    212&    58&         0&         0&         0 \\
58&    58&    179&         0&         0&         0 \\
0&         0&         0&    58&         0&         0 \\
0&         0&         0&         0&    58&         0 \\
0&         0&         0&         0&         0&    80\\
\end{bmatrix}. \label{voigt_tet_pd}
\end{equation}

The structure of the stiffness matrix is quite similar to that of orthotropic symmetry in Eq. \eqref{voigt_ortho_pd} which is not surprising as the only difference is the equality of $C_{13}, C_{23}, C_{44}, C_{55}$. Therefore it is evident that with Cauchy's relations imposed, orthotropic symmetry has 6 independent constants whereas tetragonal has 5. The error in the calibration for this particular material is found to be $3.8634\%$.

\subsubsection{Transversely Isotropic Symmetry}
Transverse isotropy is a special case of orthotropy wherein there exists a plane of isotropy resulting in 5 independent constants. It can also be thought of as being similar to tetragonal symmetry with an additional constraint on the $C_{66}$ which consequently reduces the number of independent constants from 6 to 5. Along with orthotropy, transverse isotropy has also been modeled in peridynamics, usually for composite laminates in 2D.

The symmetry group for a transversely isotropic material can be written as, 
\begin{equation}
\mathcal{S}=\Bigg\{ \bm{I}, \bm{-I}, \bm{R}(\bm{e_3}), \bm{R}\left(\frac{1}{\sqrt{a^2 + b^2}} (a\bm{e_1} + b\bm{e_2}) \right): a,b \in \mathcal{R}\Bigg\}.
\end{equation}

\noindent For example, if the $x-y$ plane is the plane of isotropy, then the 5 independent constants can be written as, 
\begin{equation}
C_{11} = C_{22}, \; C_{33}, \; C_{23} = C_{13}, \; C_{12} \; C_{55} = C_{44}, \; C_{66} = (C_{11} - C_{12})/2.   \label{trans_conditions}
\end{equation}

\noindent Similar to the trigonal symmetry, the last of Eq. \eqref{trans_conditions} along with $C_{66} = C_{12}$ in addition to other restrictions imposed by Cauchy's relations reduce the number of independent constants from 5 to 3 given by,
\begin{equation}
C_{11} = C_{22} = 3C_{12} = 3C_{66}, C_{33}, C_{23} = C_{13} = C_{44} = C_{55},
\end{equation}
and can be visualized as follows,
\begin{equation}
\begin{bmatrix} 
C_{11}&    C_{12}&    C_{13}&	0&	0&	0\\
C_{21}&    C_{22}&    C_{23}&	0&	0&	0\\
C_{31}&    C_{32}&    C_{33}&	0&	0&	0\\
0&    0&   0&	C_{44}&	0&	0\\
0&    0&   0&	0&	C_{55}&	0\\
0&    0&    0&	0&	0&	C_{66}\\
\end{bmatrix}
\rightarrow
\begin{bmatrix} 
C_{11}&    	\odot&  \Box&	0&	0&	0\\
\odot&    C_{22}&    	\Box&	0&	0&	0\\
\Box&    \Box&    C_{33}&	0&	0&	0\\
0&    0&    0&	\Box&	0&	0\\
0&    0&    0&	0&	\Box&	0\\
0&    0&    0&	0&	0&	 \odot\\
\end{bmatrix}. \label{cauchy_mat_trans}
\end{equation}

Note that Cauchy's relations render the structure of the stiffness matrix in Eq. \eqref{cauchy_mat_trans} exactly the same as that for a tetragonal material in Eq. \eqref{cauchy_mat_tet}. The only difference being that for the tetragonal case, $C_{66}$ can be defined independent of $C_{12}$. In either case, Cauchy's relations enforce $C_{66} = C_{12}$.

The reference Voigt elastic stiffness tensor for transverse isotropy (MoN for this example) is given by, 
\begin{equation}
\mathcal{\bf{\tilde{C}}}^{ref} = 
\begin{bmatrix} 
499&    177&    235&         0&         0&         0 \\
177&    499&    235&         0&         0&         0 \\
235&    235&    714&         0&         0&         0 \\
0&         0&         0&    241&         0&         0 \\
0&         0&         0&         0&    241&         0 \\
0&         0&         0&         0&         0&    161\\
\end{bmatrix}. \label{voigt_trans}
\end{equation}

\noindent The resulting bond micromoduli from the least-squares method is shown in Figure \ref{trans},

\begin{figure}[H]
	\centering
	\captionsetup{font=footnotesize}
	\captionsetup{justification=centering}	
	\begin{subfigure}[t]{0.32\textwidth}
		\centering
		\includegraphics[width=1\textwidth]{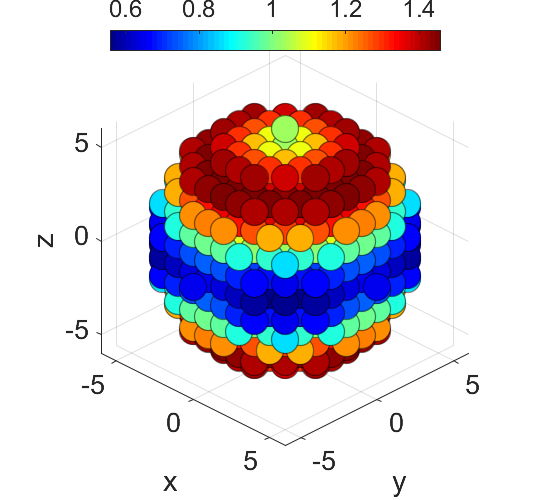}
		\caption{}
		\label{trans_1}
	\end{subfigure}%
	\begin{subfigure}[t]{0.32\textwidth}
		\centering
		\includegraphics[width=1\textwidth]{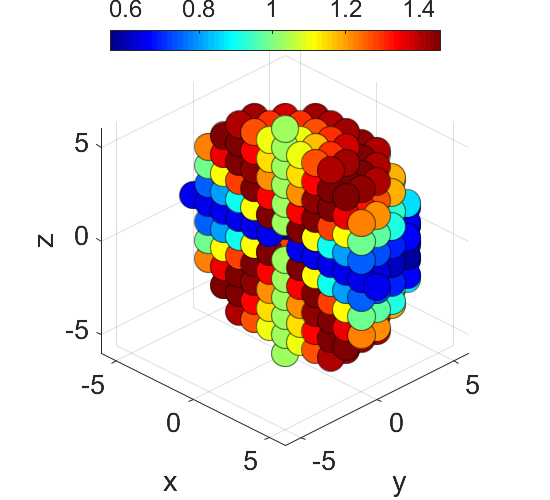}
		\caption{}
		\label{trans_2}
	\end{subfigure}
	\begin{subfigure}[t]{0.32\textwidth}
		\centering
		\includegraphics[width=1\textwidth]{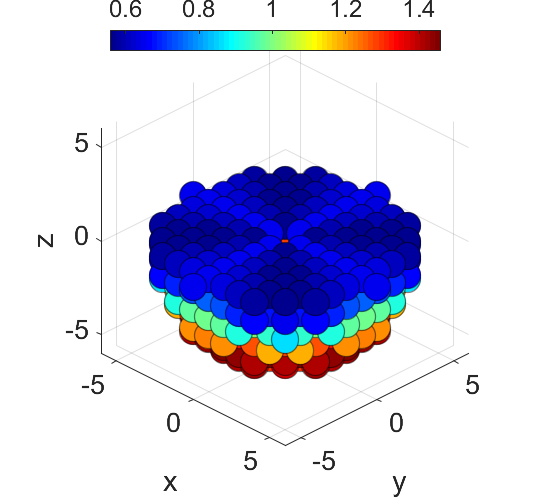}
		\caption{}
		\label{trans_3}
	\end{subfigure}
	\caption[]{(a) Complete neighborhood, and partial views of the neighborhood with normal to the cut plane at (b) $\bm{n} = -\bm{e_2}$ and (c)  $\bm{n} = \bm{e_3}$ for transversely isotropic symmetry  with horizon $\delta = 6$ and influence function \( \ifun = \frac{\delta}{\abs{\bm{\xi}}} \).}
	\label{trans}
\end{figure}

\noindent and the resulting effective peridynamics stiffness tensor is found to be,
\begin{equation}
\mathcal{\bf{\tilde{C}}}^{PD} = 
\begin{bmatrix} 
499&    166.33&    239&         0&         0&         0 \\
166.33&    499&    239&         0&         0&         0 \\
239&    239&    714&         0&         0&         0 \\
0&         0&         0&    239&         0&         0 \\
0&         0&         0&         0&    239&         0 \\
0&         0&         0&         0&         0&    166.33\\
\end{bmatrix}. \label{voigt_trans_pd}
\end{equation}

A simple visual examination of Figure \ref{trans} reveals that the bond micromoduli as a function of spatial coordinates resembles a transversely isotropic material. Higher micromoduli in the bonds with directions out of the $x-y$ plane impart the higher stiffness in the $C_{33}$ stiffness term in Eq. \eqref{voigt_trans_pd}. It is also interesting to note the difference in the distribution of bond micromoduli in plane with normal $\bm{n} = -\bm{e_2}$, which is clearly a function of the bond directions, and $\bm{n} = \bm{e_3}$ in which there is a very low variation in bond micromoduli as a function of the bond angle with respect to the $x$ axis, as shown in Figures \ref{trans_2} and \ref{trans_3}. It's easy to verify that the calibrated stiffness matrix of course still satisfies Cauchy's relations. The relative error in the calibration is found to be 1.5335\%.

\subsubsection{Cubic Symmetry}
Cubic symmetry can be generated using 9 planes of reflection symmetry, the symmetry group for which can be written as,
\begin{equation}
\mathcal{S}=\Bigg \{ \bm{I}, \bm{-I}, \bm{R}(\bm{e_1}), \bm{R}(\bm{e_2}), \bm{R}(\bm{e_3}), 
\bm{R}\left(\frac{1}{\sqrt{2}} \bm{e_1} \pm \frac{1}{\sqrt{2}} \bm{e_2}\right),
\bm{R}\left(\frac{1}{\sqrt{2}} \bm{e_2} \pm \frac{1}{\sqrt{2}} \bm{e_3}\right),
\bm{R}\left(\frac{1}{\sqrt{2}} \bm{e_1} \pm \frac{1}{\sqrt{2}} \bm{e_3}\right)
\Bigg \}.
\end{equation}

\noindent In the case of cubic symmetry, the number of independent constants are 3 given by, 
\begin{equation}
C_{11} = C_{22} = C_{33}, \; C_{12} = C_{23} = C_{13}, \; C_{44} = C_{55} = C_{66}. 
\end{equation}

\noindent Cauchy's relations reduce the number of independent constants from 3 to 2 given by,
\begin{equation}
C_{11} = C_{22} = C_{33},\; C_{12} = C_{23} = C_{13} = C_{44} = C_{55} = C_{66}. 
\end{equation}
and can be visualized as follows,
\begin{equation}
\begin{bmatrix} 
C_{11}&    C_{12}&    C_{13}&	0&	0&	0\\
C_{21}&    C_{22}&    C_{23}&	0&	0&	0\\
C_{31}&    C_{21}&    C_{33}&	0&	0&	0\\
0&    0&   0&	C_{44}&	0&	0\\
0&    0&   0&	0&	C_{55}&	0\\
0&    0&    0&	0&	0&	C_{66}\\
\end{bmatrix}
\rightarrow
\begin{bmatrix} 
C_{11}&    	\Box&  \Box&	0&	0&	0\\
\Box&    C_{22}&    	\Box&	0&	0&	0\\
\Box&    \Box&    C_{33}&	0&	0&	0\\
0&    0&    0&	\Box&	0&	0\\
0&    0&    0&	0&	\Box&	0\\
0&    0&    0&	0&	0&	 \Box\\
\end{bmatrix}. \label{cauchy_mat_cubic}
\end{equation}

An example of a material with Cubic symmetry is MgAl$_2$O$_2$ (Spinel) for which the Voigt elastic stiffness matrix is given by,
\begin{equation}
\mathcal{\bf{\tilde{C}}}^{ref} = 
\begin{bmatrix} 
252&    145&    145&         0&         0&         0 \\
145&    252&    145&         0&         0&         0 \\
145&    145&    252&         0&         0&         0 \\
0&         0&         0&    142&         0&         0 \\
0&         0&         0&         0&    142&         0 \\
0&         0&         0&         0&         0&    142\\
\end{bmatrix}. \label{voigt_cubic}
\end{equation} 

\noindent As seen in Eq. \eqref{voigt_cubic}, the material very nearly satisfies Cauchy's relations as the off-diagonal axial terms $C_{12}, C_{13}, C_{23}$ are numerically very similar to the shear terms $C_{44}, C_{55}, C_{66}$. The anisotropy index $A^{U}$ for this material is evaluated to be 1.2372.

The resulting bond micromoduli from the least-squares method is shown in Figure \ref{cubic},
\begin{figure}[H]
	\centering
	\captionsetup{font=footnotesize}
	\captionsetup{justification=centering}	
	\begin{subfigure}[t]{0.32\textwidth}
		\centering
		\includegraphics[width=1\textwidth]{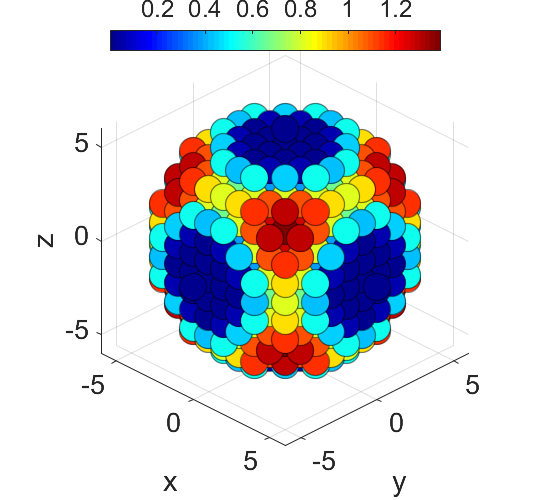}
		\caption{}
		\label{cubic_1}
	\end{subfigure}%
	\begin{subfigure}[t]{0.32\textwidth}
		\centering
		\includegraphics[width=1\textwidth]{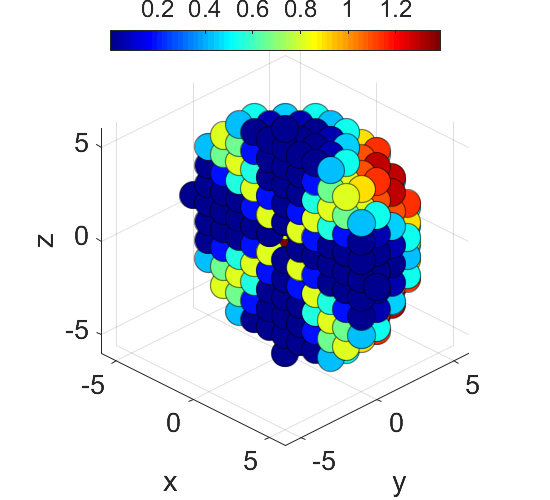}
		\caption{}
		\label{cubic_2}
	\end{subfigure}
	\begin{subfigure}[t]{0.32\textwidth}
		\centering
		\includegraphics[width=1\textwidth]{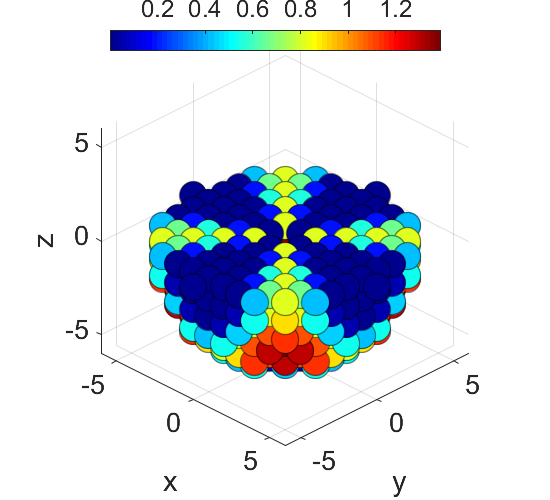}
		\caption{}
		\label{cubic_3}
	\end{subfigure}
	\caption[]{(a) Complete neighborhood, and partial views of the neighborhood with normal to the cut plane at (b) $\bm{n} = -\bm{e_2}$ and (c)  $\bm{n} = \bm{e_3}$ for cubic symmetry  with horizon $\delta = 6$ and influence function \( \ifun = \frac{\delta}{\abs{\bm{\xi}}} \).}
	\label{cubic}
\end{figure}

\noindent and the resulting effective peridynamics stiffness tensor is found to be,
\begin{equation}
\mathcal{\bf{\tilde{C}}}^{PD} = 
\begin{bmatrix} 
252&      143.5&    143.5&         0&         0&         0 \\
143.5&    252&    143.5&         0&         0&         0 \\
143.5&    143.5&    252&         0&         0&         0 \\
0&         0&         0&    143.5&         0&         0 \\
0&         0&         0&         0&    143.5&         0 \\
0&         0&         0&         0&         0&    143.5\\
\end{bmatrix}. \label{voigt_cubic_pd}
\end{equation}

The relative error in the solution for this particular material is found to be 0.8028\%.

\subsubsection{Isotropic Symmetry}
Finally, the least anisotropic symmetry - isotropy may be generated by choosing any orthogonal transformation or alternatively any reflection symmetry plane. In other words, the symmetry group for isotropy may be defined as an infinite number of orthogonal transformations. In general, for isotropic materials, the following relationships in terms of the Voigt notation will hold,
\begin{equation}
C_{11} = C_{22} = C_{33}, \;
C_{12} = C_{23} = C_{13}, \;
C_{44} = C_{55} = C_{66} = (C_{11} - C_{12})/2, \; \label{isotropic_conditions}
\end{equation} 

\noindent Isotropic symmetry reduces the number of independent elastic constants to just 2. In addition, Cauchy's relations also impose the following,
\begin{equation}
C_{44} = C_{55} = C_{66} = C_{12} = C_{13} = C_{23}, \label{isotropic_cauchy}
\end{equation} 

\noindent thereby reducing the number of elastic constants to only 1 given by,
\begin{equation}
C_{11} = C_{22} = C_{33} = 3C_{12} = 3C_{13} = 3C_{23} = 3C_{44} = 3C_{55} = 3C_{66}. \label{isotropic_cauchy_2}
\end{equation} 

\noindent The structure of the Voigt stiffness matrix can be visualized as follows,
\begin{equation}
\begin{bmatrix} 
C_{11}&    C_{12}&    C_{13}&	0&	0&	0\\
C_{21}&    C_{22}&    C_{23}&	0&	0&	0\\
C_{31}&    C_{21}&    C_{33}&	0&	0&	0\\
0&    0&   0&	C_{44}&	0&	0\\
0&    0&   0&	0&	C_{55}&	0\\
0&    0&    0&	0&	0&	C_{66}\\
\end{bmatrix}
\rightarrow
\begin{bmatrix} 
C_{11}&    	\Box&  \Box&	0&	0&	0\\
\Box&    C_{22}&    	\Box&	0&	0&	0\\
\Box&    \Box&    C_{33}&	0&	0&	0\\
0&    0&    0&	\Box&	0&	0\\
0&    0&    0&	0&	\Box&	0\\
0&    0&    0&	0&	0&	 \Box\\
\end{bmatrix}. \label{cauchy_mat_isotropic}
\end{equation}

\noindent It is easy to see from Eq. \eqref{isotropic_cauchy_2} that these conditions reduce to the well known limitation on the Poisson's ratio ($\nu = 1/4$), which of course bond-based peridynamics inherits. 
\begin{equation}
C_{11} = 3C_{12} \implies \frac{E(1-\nu)}{(1+\nu)(1-2\nu)} = \frac{3E\nu}{(1+\nu)(1-2\nu)} \implies \nu = \frac{1}{4}.
\end{equation}

The isotropic material that is chosen as an example is Pyroceram 9608 which happens to satisfy exactly Cauchy's relations \cite{trageser2019anisotropic}. The anisotropy index $A^U$ is of course, evaluated to be zero as it should be for isotropic materials. The stiffness matrix in Voigt notation is given as,
\begin{equation}
\mathcal{\bf{\tilde{C}}}^{ref} = 
\begin{bmatrix} 
103.2&    34.4&    34.4&         0&         0&         0 \\
34.4&    103.2&    34.4&         0&         0&         0 \\
34.4&    34.4&    103.2&         0&         0&         0 \\
0&         0&         0&    34.4&         0&         0 \\
0&         0&         0&         0&    34.4&         0 \\
0&         0&         0&         0&         0&    34.4 \\
\end{bmatrix}.
\end{equation} \label{voigt_iso}

Unlike evaluating a single value for the micromodulus using the analytical expression given in Eq. \eqref{silling3D}, the least-squares approach gives multiple values of bond micromoduli, even in the isotropic case. For the purposes of comparison, the micromodulus evaluated using Eq. \eqref{silling3D} is 0.2535 N/m$^6$ (with an influence function of 1) whereas the discrete calibration is a distribution bounded by 0.1726 N/m$^6$ and 0.2076 N/m$^6$ (note that the bond force has units of force per unit volume squared). 

\begin{figure}[H]
	\centering
	\captionsetup{font=footnotesize}
	\captionsetup{justification=centering}	
	\begin{subfigure}[t]{0.32\textwidth}
		\centering
		\includegraphics[width=1\textwidth]{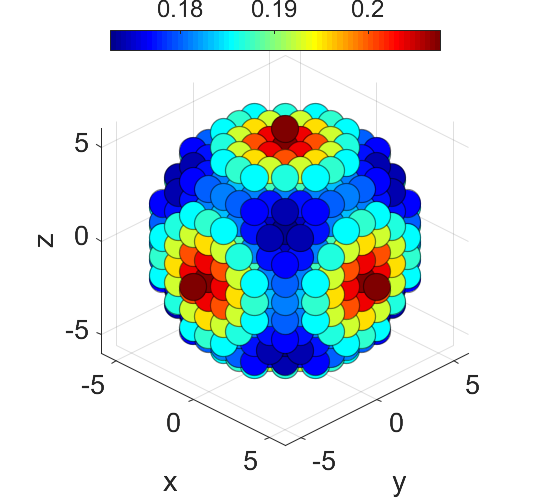}
		\caption{}
		\label{isotropic_1}
	\end{subfigure}%
	\begin{subfigure}[t]{0.32\textwidth}
		\centering
		\includegraphics[width=1\textwidth]{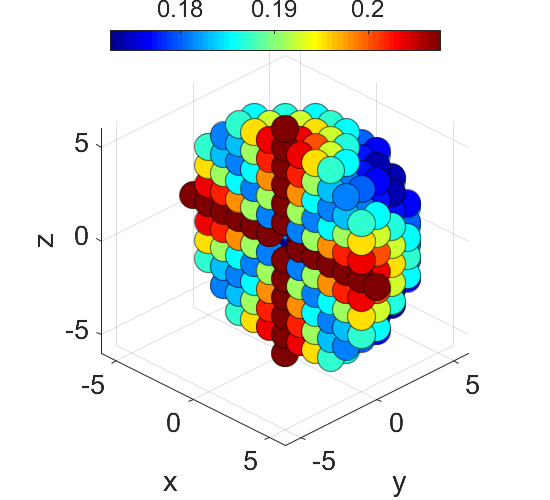}
		\caption{}
		\label{isotropic_2}
	\end{subfigure}
	\begin{subfigure}[t]{0.32\textwidth}
		\centering
		\includegraphics[width=1\textwidth]{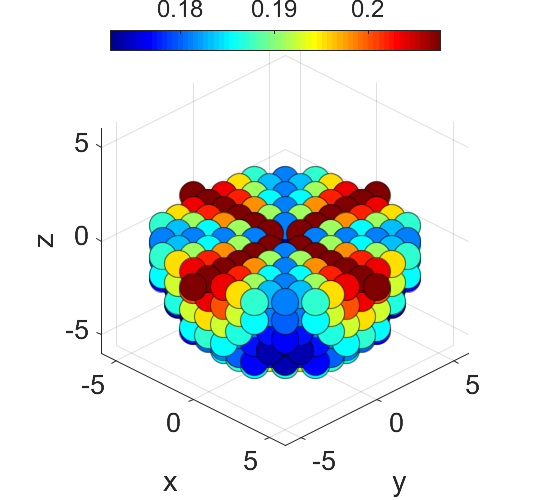}
		\caption{}
		\label{isotropic_3}
	\end{subfigure}
	\caption[]{(a) Complete neighborhood, and partial views of the neighborhood with normal to the cut plane at (b) $\bm{n} = -\bm{e_2}$ and (c)  $\bm{n} = \bm{e_3}$ for isotropic symmetry  with horizon $\delta = 6$ and influence function \( \ifun = \frac{\delta}{\abs{\bm{\xi}}} \).}
	\label{isotropic}
\end{figure}

The bond micromoduli thus obtained are shown graphically in Figure \ref{isotropic}. The bond micromoduli do depend on bond orientation with respect to material axes however ultimately results in an effective isotropic material. The effective peridynamic stiffness tensor obtained after calibration using the least-squares method in Eq. \eqref{disc_stiff_tens} is found to be exactly equal to the reference stiffness tensor. The relative error in the solution is effectively zero ($e = 2.7225\times10^{-11}$). 

\begin{figure}[H]
	\centering
	\captionsetup{font=footnotesize}
	\captionsetup{justification=centering}	
	\begin{subfigure}[t]{0.32\textwidth}
		\centering
		\includegraphics[width=1\textwidth]{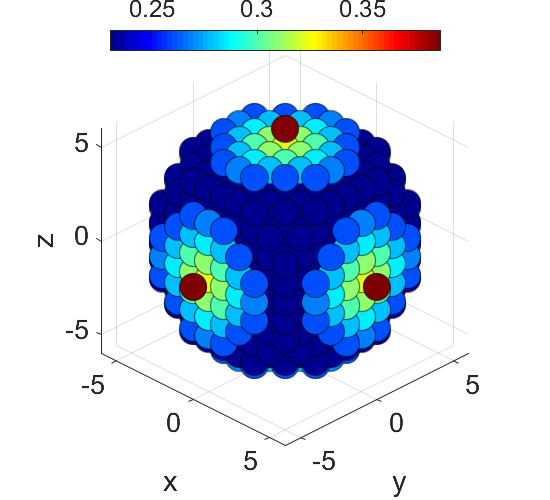}
		\caption{}
		\label{iso_const_1}
	\end{subfigure}%
	\begin{subfigure}[t]{0.32\textwidth}
		\centering
		\includegraphics[width=1\textwidth]{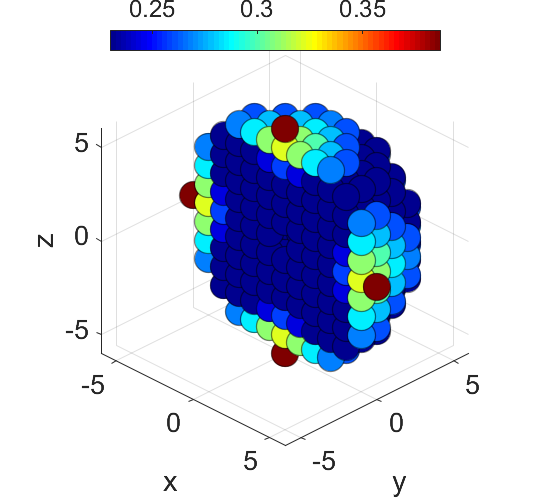}
		\caption{}
		\label{iso_const_2}
	\end{subfigure}
	\begin{subfigure}[t]{0.32\textwidth}
		\centering
		\includegraphics[width=1\textwidth]{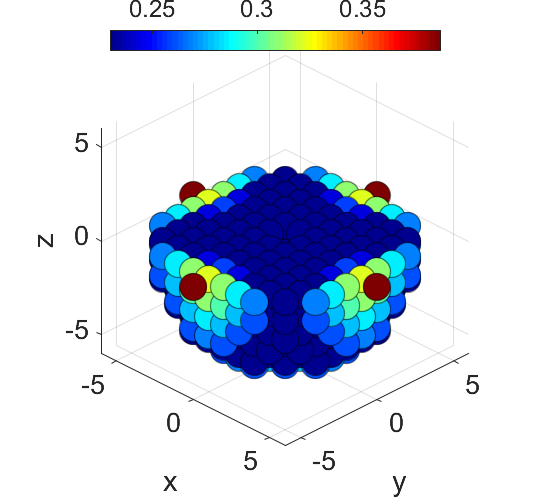}
		\caption{}
		\label{iso_const_3}
	\end{subfigure}
	\caption{(a) Complete neighborhood, and partial views of the neighborhood with normal to the cut plane at (b) $\bm{n} = -\bm{e_2}$ and (c)  $\bm{n} = \bm{e_3}$ for isotropic symmetry with horizon $\delta = 6$ and influence function $\protect\ifun = 1$. An additional lower bound on the solution of 0.23 N/m$^6$ is applied.}
	\label{isotropic_constrained}
\end{figure}

One may find a non-constant micromoduli distribution unusual and counter-intuitive, understandably, as the most commonly adopted approach is to use a constant micromodulus. However, it can be shown that a nearly constant micromodulus distribution can be recovered using the proposed method as well. This can be done by using a well informed choice of bounds in the quadratic program defined in Eqs. \eqref{min_1},  \eqref{min_2} and \eqref{min_3}. The analytical value of the micromodulus using an influence function, $\ifun = 1$ is known to be 0.2535 N/m$^6$. Therefore, choosing a lower bound of 0.23 N/m$^6$ (after some numerical experimentation) results in a solution that is given in Figure \ref{isotropic_constrained}. It is quite evident that the micromoduli are nearly constant within the horizon except for some bonds near the edge of the spherical horizon. The mean value of this distribution is found to be 0.2435 N/m$^6$ which agrees quite well with the analytical value of 0.2535 N/m$^6$. Note that, a perfect match is not expected since (by a rough estimation) the volume of the full spherical neighborhood is approximately $905m^3$, compared to a value of $925m^3$ for the discrete neighborhood. Accounting for the slight difference in volume, the analytical value can be scaled down, according to the difference in volume, to arrive at a true analytical estimate of 0.2480 N/m$^6$ which is in excellent agreement with the mean value of 0.2435 N/m$^6$ for the discrete distribution.

It is important to note that the calibrated peridynamic stiffness tensor always satisfies Cauchy's relations regardless of the material symmetry. This implies that if the reference stiffness matrix that is input to the proposed least-squares method satisfies Cauchy's relations a priori, then the relative error of calibration is effectively zero (close to machine precision) as seen for the isotropic case. This can be done by modifying the material's stiffness matrix appropriately using an analytical approach such as that given in \cite{trageser2019anisotropic}. The fact that the relative error is zero also indicates that volume correction is accounted for, not surprising since a discretized neighborhood is being used.
\begin{table}[H]
	\small
	\resizebox{\textwidth}{!}{%
		\begin{tabular}{|c|c|c|c|c|c|c|}
			\hline
			\textbf{Symmetry}                                                 & \textbf{\begin{tabular}[c]{@{}c@{}}Constants\\ (Traditional)\end{tabular}} & \textbf{\begin{tabular}[c]{@{}c@{}}Constants\\ (Cauchy's rel. imposed)\end{tabular}} & \textbf{Example} & \textbf{A$^U$} & \textbf{\begin{tabular}[c]{@{}c@{}}Error (\%)\\ (Original)\end{tabular}} & \textbf{\begin{tabular}[c]{@{}c@{}}Error (\%)\\ (Cauchy's rel. imposed)\end{tabular}} \\ \hline
			Triclinic 	& 21 & 15  & KIO$_3$           	& 0.1965      & 3.1873  	& 8.1291e-11	\\ \hline
			Monoclinic 	& 13 & 9 & CoTeO$_4$           	& 28.3169     & 4.988   	& 1.1187e-13    \\ \hline
			Orthotropic & 9  & 6 & Te$_2$W             	& 54.0623     & 5.0958  	& 9.2958e-14    \\ \hline
			Trigonal & 6 & 4 & Ta$_2$C             		& 0.5875      & 2.632   	& 8.1990e-14    \\ \hline
			Tetragonal & 6 & 4 & Si               		& 0.1231      & 3.8634  	& 1.1168e-11    \\ \hline
			Trans. Isotropic & 5 & 3 & MoN              & 0.2551      & 1.5335  	& 6.9451e-11    \\ \hline
			Cubic & 3 & 2  & MgAl$_2$O$_2$ (Spinel) 	& 1.2372      & 0.8028  	& 1.5723e-13    \\ \hline
			Isotropic & 2 & 1  & Pyroceram 9608    		& 0           & 2.7225e-11 	& 2.7225e-11    \\ \hline
	\end{tabular}}
	\normalsize
	\caption{Summary of results for all eight symmetries tested.}
	\label{table_summary}
\end{table} 

Table \ref{table_summary} and Figure \ref{all_symmetries} summarize results obtained from all eight symmetries. The maximum error in the calibration procedure was found to occur for the orthotropic symmetry, where as the lowest error was found in the case of isotropy. The last column also shows the relative error when the reference material stiffness matrix is modified such that Cauchy's relations are satisfied a priori. Not surprisingly, it is found that the errors in calibration are effectively zero. Most importantly, no evidence of correlation between the universal anisotropy constant and the error is found, or in other words, the calibration error was found to depend on how closely the material satisfies Cauchy's relations but not on the material symmetry itself.

\begin{figure}[H]
	\centering
	\captionsetup{font=footnotesize}
	\captionsetup{justification=centering}
	\includegraphics[width=1\textwidth, trim={2cm 2.5cm 2cm 2cm},clip]{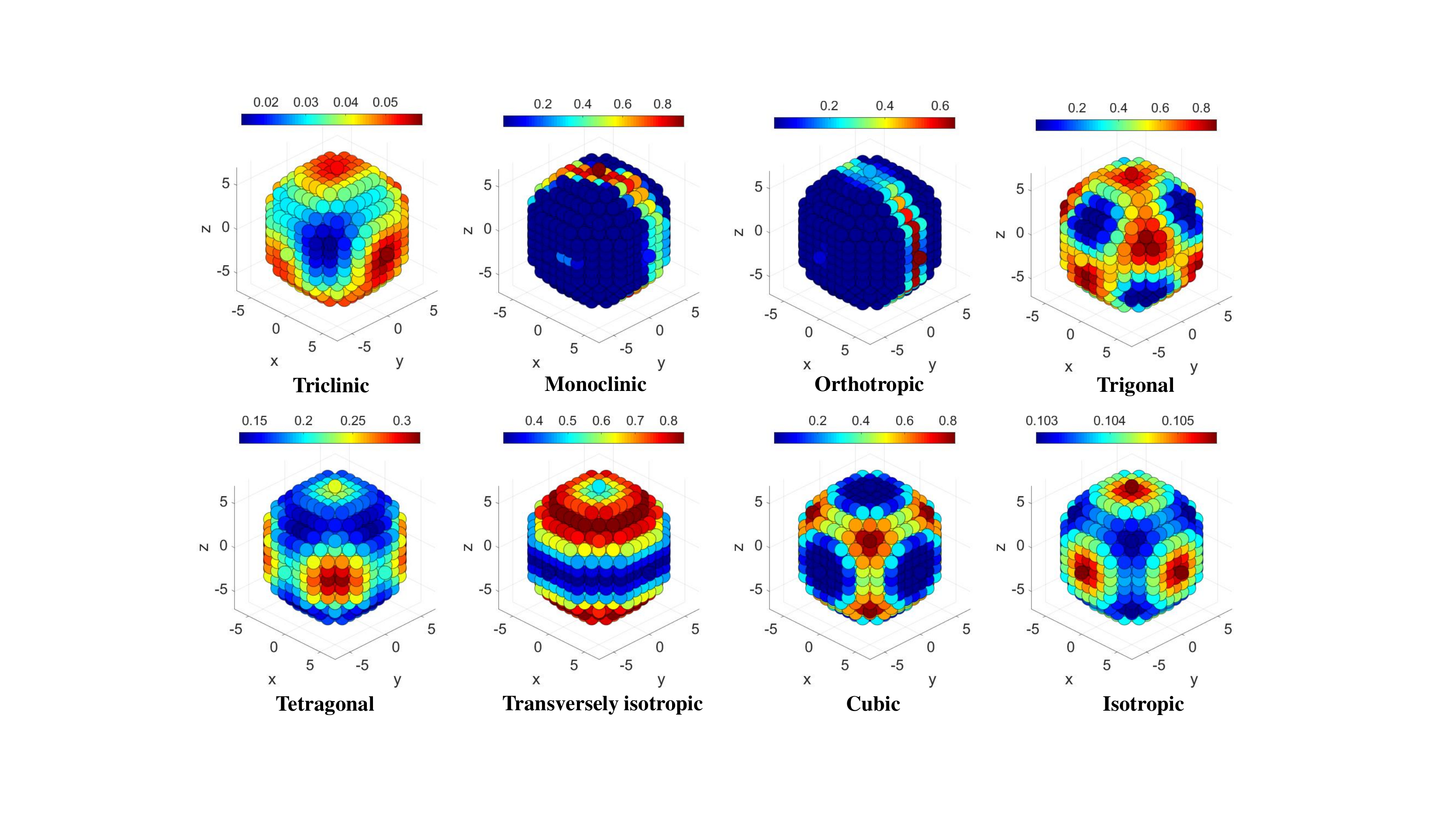}
	\caption{All eight symmetries using a horizon ratio of 6 and an inverse influence function.}
	\label{all_symmetries}
\end{figure}

\subsection{Parametric Studies}
Since the method described in this work is a numerical method, various parametric studies are conducted to demonstrate the generality and robustness of the method. While combinations of all parameters are endless, a few of practical significance are investigated, namely the choice of horizon ratio, horizon shape, influence function and rotation of the lattice or the neighborhood. Numerical experiments with lattice rotations also serve as verification that the proposed method presents solutions that are meaningful and satisfy conditions of the material's symmetry group. 

\subsubsection{Effect of Horizon Ratio}
Naturally, one of questions that is of practical importance is how the horizon ratio affects the calibration procedure, i.e. does the number of bonds in the neighborhood affect the quality of the solution?
\begin{figure}[H]
	\centering
	\captionsetup{font=footnotesize}
	\captionsetup{justification=centering}
	\includegraphics[width=1\textwidth, trim={1cm 0cm 1cm 0cm},clip]{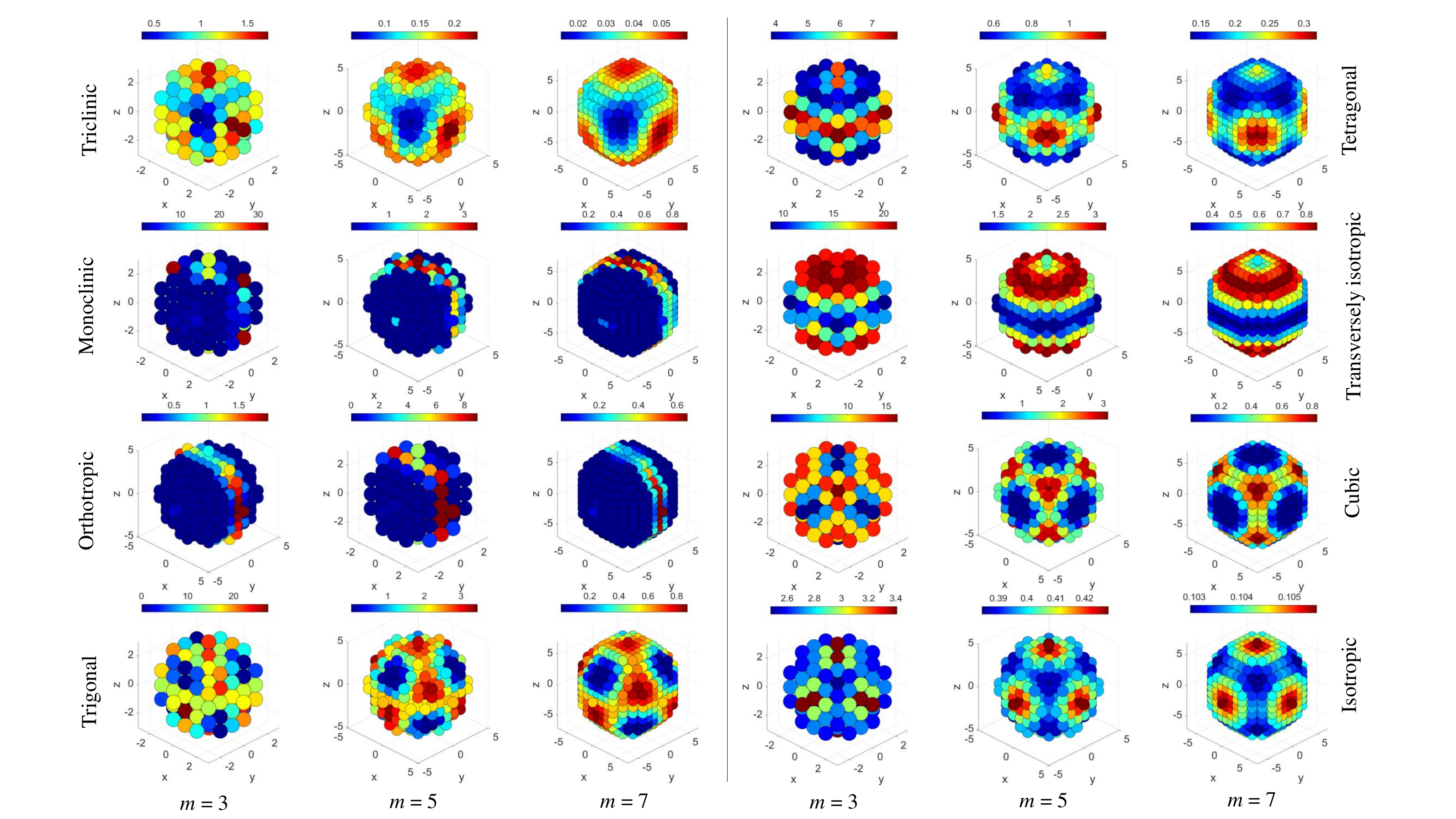}
	\caption{All eight symmetries using example materials from section \ref{section_8_symm} with horizon ratio $m$ of 3, 5 and 7.}
	\label{horizon_ratio}
\end{figure}

In general it appears, from numerical experimentation, that the calibration procedure is unaffected by the size of the neighborhood, provided the horizon is not impractically small. Figure \ref{horizon_ratio} shows calibrated neighborhoods assuming horizon ratios of 3, 5 and 7 for all eight material symmetries using the same example materials as in section \ref{section_8_symm}. Naturally the magnitude of a typical bond micromodulus for a particular material changes with horizon size, decreasing with an increase in horizon in general due to a larger number of bonds, but the overall pattern simply from visual inspection is the same. Moreover, it is observed that the error in the calibration is unaffected for most reasonable values of the horizon ratio (including all the cases shown in Figure \ref{horizon_ratio}). 

For a horizon as small as 2 times the lattice spacing, a convergent solution is found for most cases, some with an increase in the relative error. For example, the error remains exactly the same when using a horizon of 2 for isotropic, cubic, transversely isotropic, tetragonal and orthotropic symmetries but the error increases to 6.732\% for trigonal and 3.9036\% for triclinic symmetry, where as a converged solution is not found for the monoclinic case. Notice that the anisotropy index for the monoclinic material in these cases is not as high as the orthotropic material, however a solution is still found for the orthotropic material. Since the monoclinic material has only a single plane of symmetry, it appears that there simply are not enough bonds that can contribute to the desired anisotropic elasticity tensor.

For all practical purposes since generally a horizon of 3 or more is used, the proposed numerical method is found to be independent of horizon size.

\subsection{Effect of Neighborhood Shape}
While a circular neighborhood in 2D and spherical neighborhood in 3D are the most common because of the ease of finding analytical expressions for micromoduli in polar or spherical coordinates respectively, the choice of neighborhood can in fact be arbitrary. For example, Ahadi and Krockmal recently used an ellipsoidal influence function with a elastic-plastic ordinary state based model \cite{ahadi2018anisotropic}. In a recent paper, Li et al. used a symmetric square shaped horizon for the analysis of functionally graded materials \cite{LI2020}. In this parametric study, three different types of neighborhoods are considered namely cubic, cuboid and ellisoidal in addition to the spherical neighborhood considered in the results section. The cubic neighborhood is assumed to have equal sides of $2\delta$, the cuboid neighborhood has sides of $\frac{8\delta}{6}, \frac{10\delta}{6}, 2\delta$ in the $x,y,z$ directions respectively and the ellipsoidal neighborhood has semi-axes of $\frac{4\delta}{6}, \frac{5\delta}{6}, \delta$ in the $x,y,z$ directions respectively, where $\delta=6$. The choice of edge lengths and semi-axes lengths for cuboid and ellipsoidal shapes are arbitrary and the same procedure could be applied to any reasonably sized cuboids or ellipsoids.

\begin{figure}[H]
	\centering
	\captionsetup{font=footnotesize}
	\captionsetup{justification=centering}
	\includegraphics[width=1\textwidth, trim={1cm 0cm 1cm 0cm},clip]{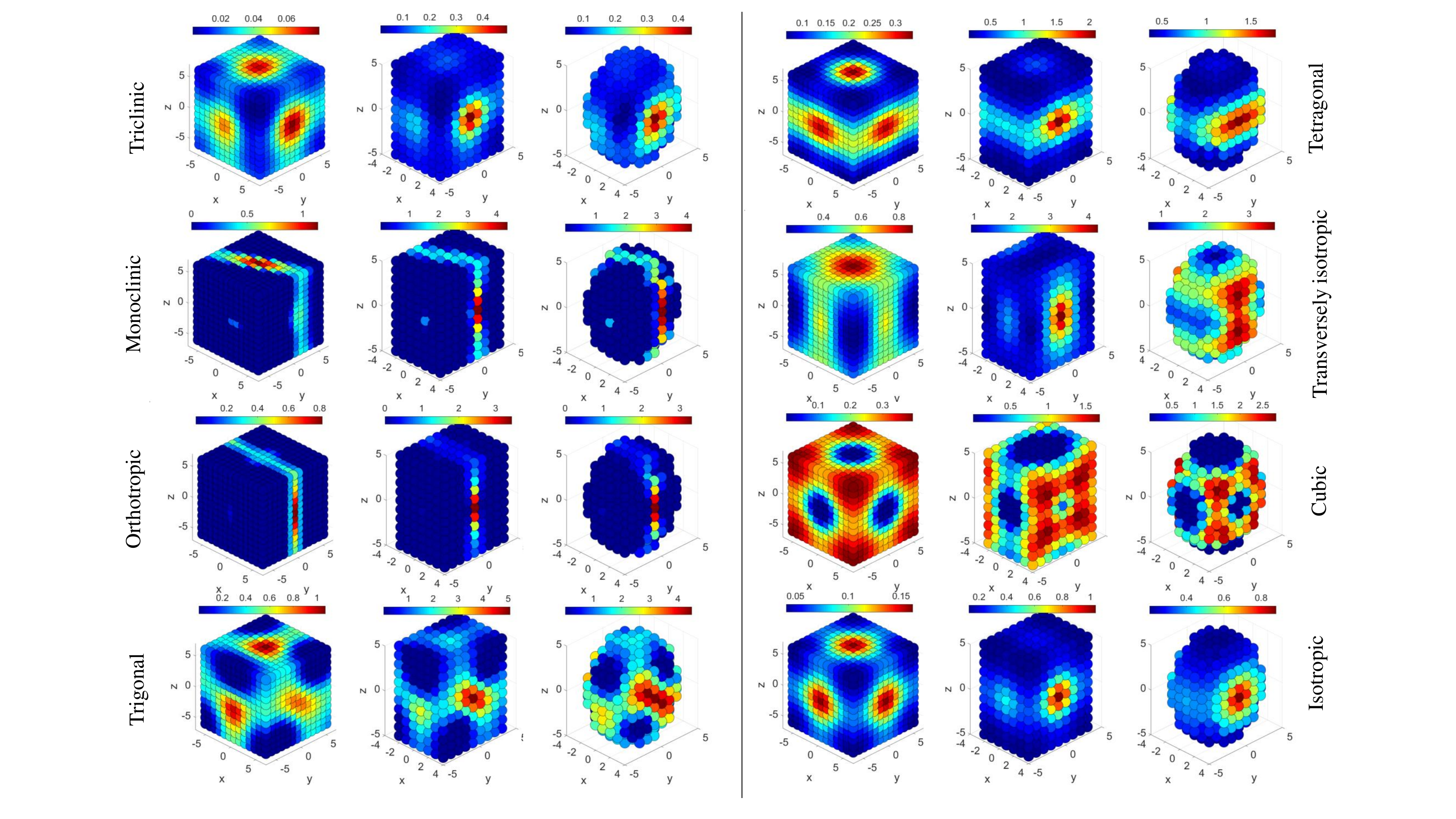}
	\caption{All eight symmetries using example materials from section \ref{section_8_symm} with a cubic, cuboid and ellipsoid neighborhoods.}
	\label{horizon_shape}
\end{figure}

Firstly, for all cases tested, the distribution of micromoduli could be successfully found using the least-squares method with no change in the errors from those given in Table \ref{table_summary}. There appear to be some interesting consequences for the micromoduli based on the type of neighborhood chosen. In most cases, the distribution of micromoduli for the cubic neighborhood is similar to it's spherical counterpart. However, in some cases like tetragonal symmetry, the distribution is notably different. While the bonds with the highest micromoduli are found to lie in the in the $x-y$ plane at angles $\pi/8, 3\pi/8, 5\pi/8$ and $7\pi/8$ using a spherical neighborhood, they appear to be at angles of $0, \pi/4, \pi/2$ and $3\pi/4$ when using a cubic neighborhood. While this seems counterintuitive at first glance, note that the longest bonds in the $x-y$ plane in the cubic neighborhood are of length $6\sqrt{2}\delta$, they are only of length $6\delta$ in the spherical case. Quite naturally, there are more bonds in the $x-y$ plane for the cubic case that can contribute to the stiffness matrix. Note also that the magnitude of micromoduli for the cubic neighborhood are slightly smaller than the spherical neighborhood. The same principle applies for the cuboid and ellipsoid neighborhoods as well - since the neighborhood contains more bonds in the $z$ direction, the magnitude of micromoduli for those bonds is lower than those oriented in the other two directions. Nevertheless, all four neighborhoods lead to same effective peridynamic stiffness matrix. Similar behavior can be seen clearly for the transversely isotropic material as well, but it is not so obvious for some others like the trigonal or cubic symmetries. In effect, it is demonstrated that the proposed method is robust enough to handle a wide range of discretized neighborhoods.

\subsubsection{Effect of Influence Function}
Another issue of practical significance is the choice of influence function $\ifun$. The influence function can be chosen to weight the contribution of bonds based on the bond lengths. As Seleson and Parks note, the choice of influence function can permit a rich spectrum of dynamic behavior in non-local models \cite{seleson2011role}. The examples in section \ref{section_8_symm} used an inverse weight function of $\delta/\abs{\bm{\xi}}$. In addition to the inverse weight function, three other choices are considered - a constant weight $ \protect\ifun=1 $, a hat function $\ifun=1-\frac{{|\bm{\xi}|}}{\delta}$ and a power law type function $ \protect\ifun=1-\protect\left( \frac{|\bm{\xi}|}{\delta} \right)^{-\frac{3}{2}} $, for each of the example materials. While the inverse influence function is singular at the center for zero bond lengths, decreasing to a minimum of unity at the horizon, the hat and the power law functions start at unity at the center and decrease to zero at the horizon. While the choice of influence function combined with a choice of the neighborhood shape can yield endless combinations, for this parametric study a spherical horizon is chosen. The results for all cases are given in Figure \ref{influence_function}, the neighborhoods are sectioned so as to be able to observe the bond distributions better.
\begin{figure}[H]
	\centering
	\captionsetup{font=footnotesize}
	\captionsetup{justification=centering}
	\includegraphics[width=1\textwidth, trim={1.5cm 0cm 1.5cm 0cm},clip]{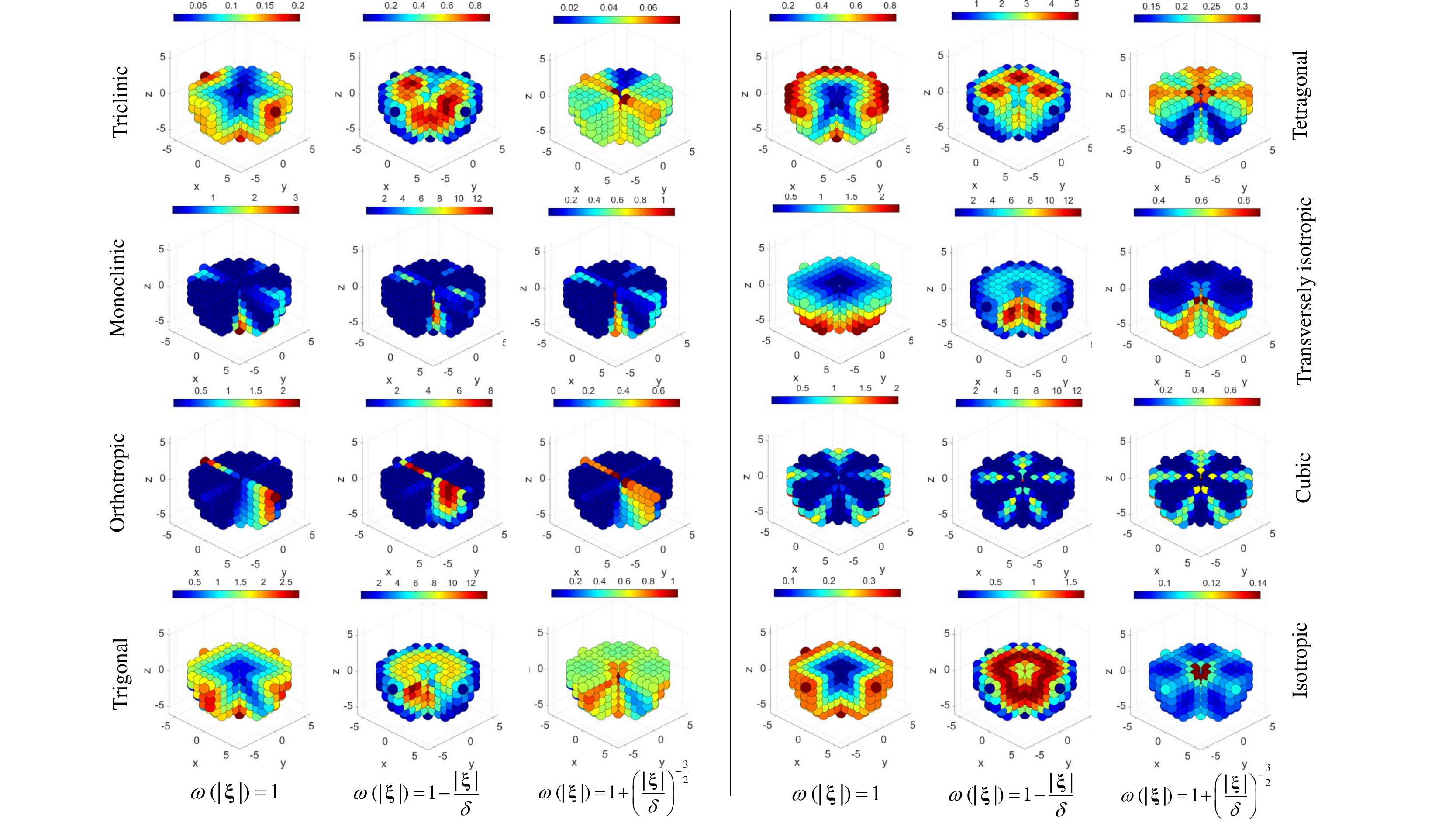}
	\caption{All eight symmetries using example materials from section \ref{section_8_symm} with influence functions $ \protect\ifun=1 $, $ \protect\ifun=1-\protect\frac{{|\bm{\xi}|}}{\delta} $ and  $ \protect\ifun=1-\protect\left( \frac{|\bm{\xi}|}{\delta} \right)^{-\frac{3}{2}} $ }
	\label{influence_function}
\end{figure}

In general, the bond micromoduli appear to be weighted quite intuitively in accordance with the nature of the influence function. This is more obvious with certain symmetries than others. For example in the case of tetragonal symmetry, the highest micromoduli are found for bonds that are longer and close to the horizon at angles of $\pi/8, 3\pi/8, 5\pi/8$ and $7\pi/8$ in the $x-y$ plane for a constant influence function of 1. With a hat function that decreases from 1 for $\abs{\bm{\xi}} = 0$ to 0 at $\abs{\bm{\xi}} = \delta$, the bonds with the highest micromoduli shift towards the center of the neighborhood, roughly for bond lengths of $2\sqrt{2}\Delta$. For the power law, this shifts further inwards to $\sqrt{2}\Delta$. Similar behavior can be observed very clearly in the isotropic case as well where the bonds with the highest micromoduli shift from $6\Delta$ to $2\Delta$ and further to $\Delta$. 

In all, this parametric study demonstrates that the least-squares method of calibration if general enough to successfully evaluate the distribution of micromoduli for a wide range of influence functions.

\subsubsection{Effect of Lattice Rotation}
Lattice\footnote{The neighborhood is also interchangeably referred to as the `lattice'.} rotations can be of two types, arbitrary rotations of the lattice that may not fall within the symmetry group of a material and specific rotations that do. In the first case, the peridynamic elasticity tensor evaluated in the direction of the rotated lattice, in general, may not be equal to that in the unrotated lattice. However, both lattices should ideally result in the same elasticity tensor in any fixed coordinate system. In the second case where the rotation imposed on the lattice falls within the symmetry group of the material, the first condition should be met in addition to the fact that the elasticity tensors evaluated in the original and rotated configurations should be equal. The following test is proposed to verify that the numerical method gives accurate micromoduli distributions under both types of lattice rotations.

\begin{figure}[H]
	\centering
	\captionsetup{font=footnotesize}
	\captionsetup{justification=centering}
	\includegraphics[width=0.5\textwidth, trim={4cm 3cm 4cm 3cm},clip]{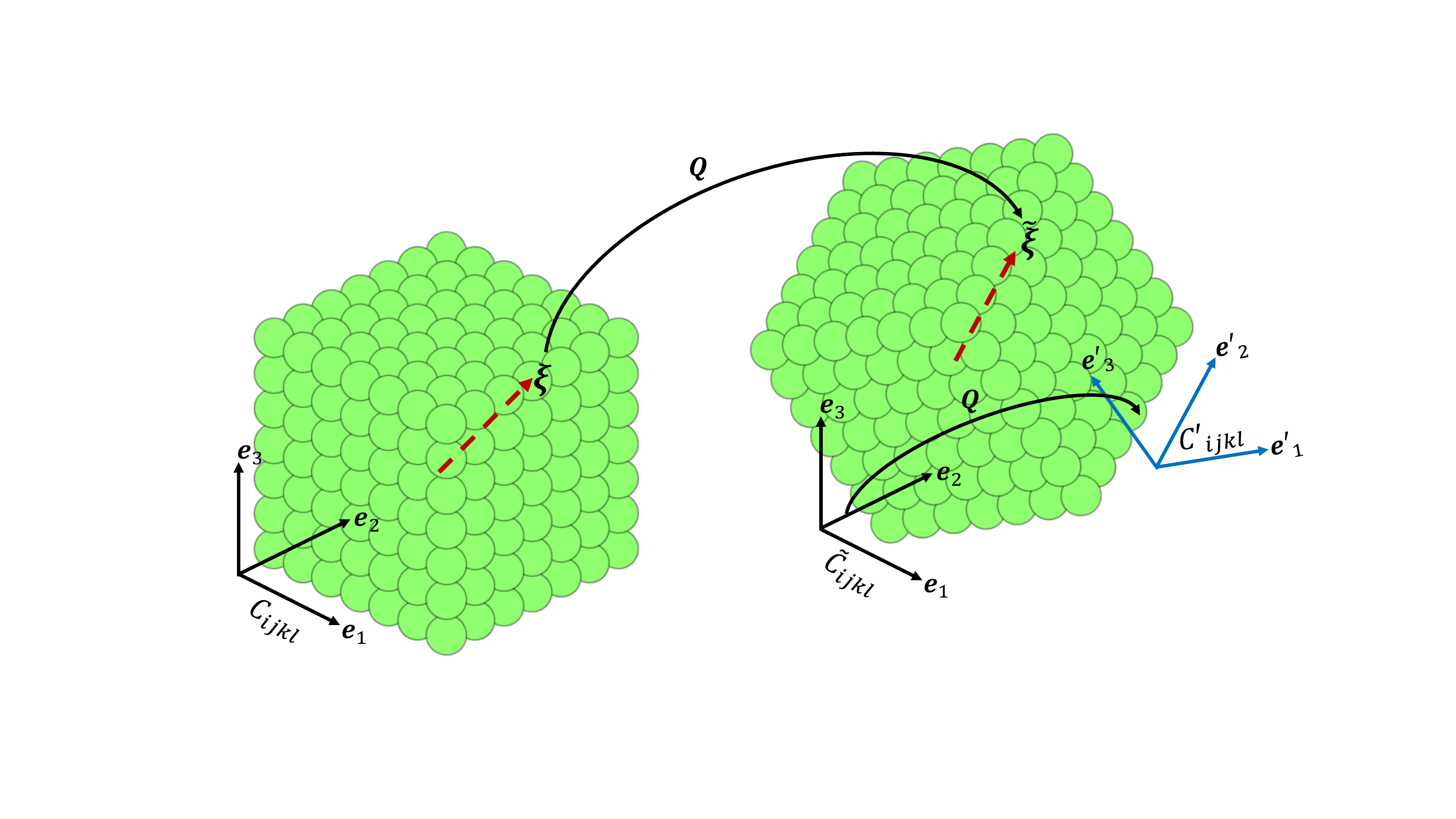}
	\caption{Schematic of the original unrotated configuration of the neighborhood/lattice and the rotated configuration along with the original unrotated coordinate basis $\{\bm{e_1}, \bm{e_2}, \bm{e_3}\}$ and the coordinate basis $\{\bm{e_1}', \bm{e_2}', \bm{e_3}'\}$ aligned with the rotated lattice. }
	\label{schematic_rotation}
\end{figure}

Assume that a peridynamic neighborhood/lattice is calibrated to some elasticity tensor where both the lattice and the elasticity tensor are in an orthogonal coordinate system $\{\bm{e_1}, \bm{e_2}, \bm{e_3}\}$ and the lattice is aligned with the coordinate system such that,
\begin{equation}
\mathbb{C}^{PD}_{ijkl} = \half \int_{H_{\bm{x}}} \ifun c\left(\bm{\xi}\right)\; \frac{\xi_i \xi_j \xi_k \xi_l}{ \blen^3 } \dvol. \label{ver_1}
\end{equation} 

\noindent Let us assume that another lattice is calibrated to the same reference elasticity tensor but the lattice is in a different orientation not necessary aligned with $\{\bm{e_1}, \bm{e_2}, \bm{e_3}\}$ such that the peridynamic stiffness tensor is then expressed as,
\begin{equation}
\tilde{\mathbb{C}}^{PD}_{ijkl} = \half \int_{H_{\bm{x}}} \ifun \tilde{c}\left(\tilde{\bm{\xi}}\right)\; \frac{\tilde{\xi}_i \tilde{\xi}_j \tilde{\xi}_k \tilde{\xi}_l}{ \blen^3 } \dvol. \footnote{For the sake of simplicity, the notiation in the influence function is left unchanged as the influence function depends only on the length of bond which is invariant with respect to the change of coordinate basis.} \label{ver_2}
\end{equation} 

\noindent The lattice in both  Eq. \eqref{ver_1} and Eq. \eqref{ver_2} are assumed to be exactly the same, in terms of the horizon, number of bonds etc., except for the orientation such that every bond satisfies the transformation $\tilde{\xi}_i = Q_{ij}\xi_j$. Then the condition for the numerical results from the least-squares method to be invariant with lattice rotations would be,
\begin{equation}
\mathbb{C}^{PD}_{ijkl} = \tilde{\mathbb{C}}^{PD}_{ijkl} \label{ver_3}
\end{equation}

Let us also assume that the proposed least-squares method is able to calibrate the bond micromoduli in both cases accurately such that Eq. \eqref{ver_3} is assumed to be satisfied. Then,
\begin{equation}
\mathbb{C}^{PD}_{ijkl} = \half \int_{H_{\bm{x}}} \ifun c\left(\bm{\xi}\right)\; \frac{\xi_i \xi_j \xi_k \xi_l}{ \blen^3 } \dvol = 
\tilde{\mathbb{C}}^{PD}_{ijkl} = \half \int_{H_{\bm{x}}} \ifun \tilde{c}\left(\tilde{\bm{\xi}}\right)\; \frac{\tilde{\xi}_i \tilde{\xi}_j \tilde{\xi}_k \tilde{\xi}_l}{ \blen^3 } \dvol. \label{ver_4}
\end{equation} 

\noindent Note that $\mathbb{C}^{PD}_{ijkl}$ and $\tilde{\mathbb{C}}^{PD}_{ijkl}$ are defined in the same coordinate system $\{\bm{e_1}, \bm{e_2}, \bm{e_3}\}$ even though the distribution of micromoduli and bond vectors are different. The peridynamic elasticity tensor can also be defined in the rotated coordinate system $\{\bm{e_1}', \bm{e_2}', \bm{e_3}'\}$ aligned with the lattice in Eq. \eqref{ver_2} such that $e'_i=Q_{ij}e_j$. Then the peridynamic stiffness tensor in the rotated coordinate system can be written as,
\begin{equation}
\mathbb{C}'^{PD}_{ijkl} = \half \int_{H_{\bm{x}}} \ifun \tilde{c}\left(\tilde{\bm{\xi}}\right)\; \frac{\xi'_i \xi'_j \xi'_k \xi'_l}{ \blen^3 } \dvol. \label{ver_5}
\end{equation} 

\noindent Note that the bond distribution $\tilde{c}\left(\tilde{\bm{\xi}}\right)$ of the rotated lattice remains the same, only the stiffness tensor is expressed in the rotated coordinate system. The primed bond vectors $\bm{\xi}'$ and  $\tilde{\bm{\xi}}$ are essentially the same vectors in different coordinate systems such that the transformation $\tilde{\xi}_i=Q_{ij}\xi'_j$ and the transformation $\tilde{\xi}_i=Q_{ij}\xi_j$ can be used to write, 
\begin{equation}
\mathbb{C}'^{PD}_{ijkl} = \half \int_{H_{\bm{x}}} \ifun \tilde{c}\left(\tilde{\bm{\xi}}\right)\; \frac{\xi_i \xi_j \xi_k \xi_l}{ \blen^3 } \dvol, \label{ver_6}
\end{equation}

\noindent Equation Eq. \eqref{ver_6} retains the micromoduli distribution $\tilde{c}\left(\tilde{\bm{\xi}}\right)$ from Eq. \eqref{ver_5}. Now, if the following relation for the micromoduli distribution holds,
\begin{equation}
\tilde{c}\left(\tilde{\bm{\xi}}\right) = c\left(\bm{\xi}\right), \label{ver_7}
\end{equation}

\noindent then Eq. \eqref{ver_6} can be re-written as,
\begin{equation}
\mathbb{C}'^{PD}_{ijkl} = \mathbb{C}^{PD}_{ijkl} \label{ver_8},
\end{equation}

\noindent which implies that the transformation $Q_{ij}$ must lie in the symmetry group for the material represented by $\mathbb{C}^{PD}_{ijkl}$. In other words, if the bond distribution of an unrotated and rotated lattice remains the same, then that transformation must lie in the symmetry group of that material. Therefore, this simple test can also be used to verify the numerical results. Although there are infinitely many combinations of symmetries and lattice rotations that can be tested, we choose three cases to demonstrate. For all cases, a cubic lattice of semi edge length of 3 is chosen resulting in a total of 342 bonds.

The first symmetry chosen is triclinic, the most anisotropic with no  planes of symmetry. The most elementary of transformations are chosen - reflections about the $\bm{e}_1, \bm{e}_2, \bm{e}_3$ axes. 
\begin{figure}[H]
	\centering
	\captionsetup{font=footnotesize}
	\captionsetup{justification=centering}
	\includegraphics[width=1\textwidth, trim={4cm 3cm 4cm 3cm},clip]{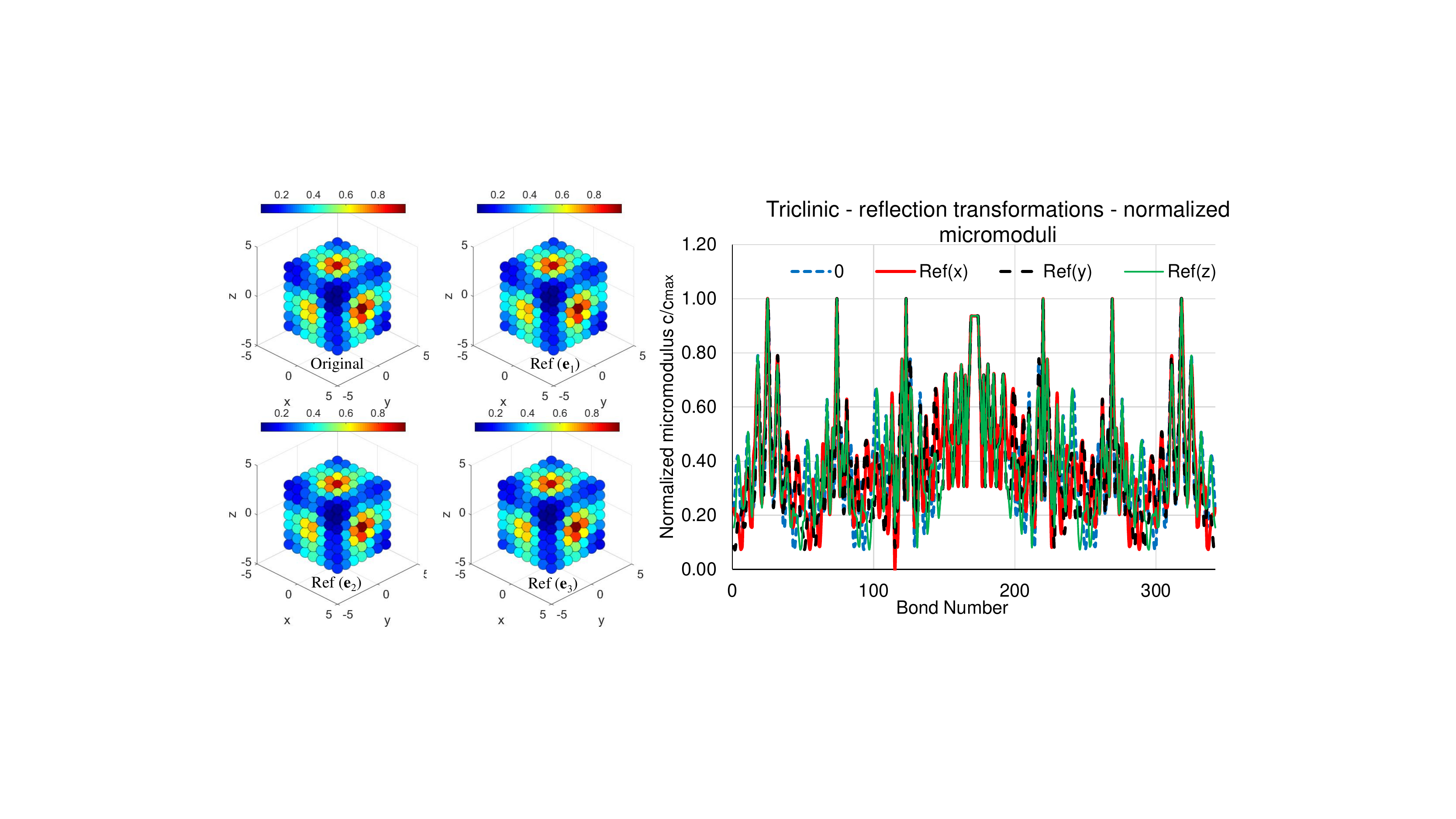}
	\caption{Reflection transformations about the $\bm{e}_1, \bm{e}_2, \bm{e}_3$ axes (left) and the distribution of normalized micromoduli $c/c_{max}$ where $c_{max}$ is the maximum value of the micromodulus in the original configuration(right) for all three transformations along with the original unrotated configuration for triclinic symmetry.}
	\label{tric_reflection}
\end{figure}

Figure \ref{tric_reflection} shows the original configuration along with the three reflection transformations imposed. In addition, the figure also shows a plot of the normalized micromoduli versus the bond number for all four cases (original unrotated configuration included). In all cases, the calibration error of 3.1873\% obtained in section \ref{section_8_symm} is unchanged. This is not surprising as the transformations in this case keep the cubic lattice orthogonal to the coordinate system, therefore the distribution of the micromoduli with respect to \textit{spatial coordinates} is exactly the same in all cases as seen in Figure \ref{tric_reflection} (left). However, the micromoduli corresponding to specific bonds in all cases are different as seen in the plot in Figure \ref{tric_reflection} (right), meaning that the transformations do not fall under the symmetry group of the material according to Eq. \eqref{ver_7} and Eq. \eqref{ver_8}.

Next, a series of rotations given by the rotation transformation $\bm{R}$ are imposed such that $\bm{R} = \bm{R}_x\bm{R}_y\bm{R}_z$, where $\bm{R}_x$, $\bm{R}_y$ and $\bm{R}_z$ are equi-angle rotations of $\theta$ about the $\bm{e}_1, \bm{e}_2, \bm{e}_3$ axes given by,
\begin{equation}
\bm{R}_x=
\begin{bmatrix} 
1 & 0 & 0 \\
0 & \cos\theta & -\sin\theta \\
0 & \sin\theta & \cos\theta \\
\end{bmatrix},
\bm{R}_y=
\begin{bmatrix} 
\cos\theta & 0 & -\sin\theta \\
0 & 1 & 0 \\
\sin\theta & 0 & \cos\theta \\
\end{bmatrix},
\bm{R}_z=
\begin{bmatrix} 
\cos\theta & -\sin\theta &  0\\
\sin\theta & \cos\theta & 0 \\
0 & 0 & 1 \\
\end{bmatrix}.
\end{equation}

\begin{figure}[H]
	\centering
	\captionsetup{font=footnotesize}
	\captionsetup{justification=centering}
	\includegraphics[width=1\textwidth, trim={2cm 3cm 2cm 3cm},clip]{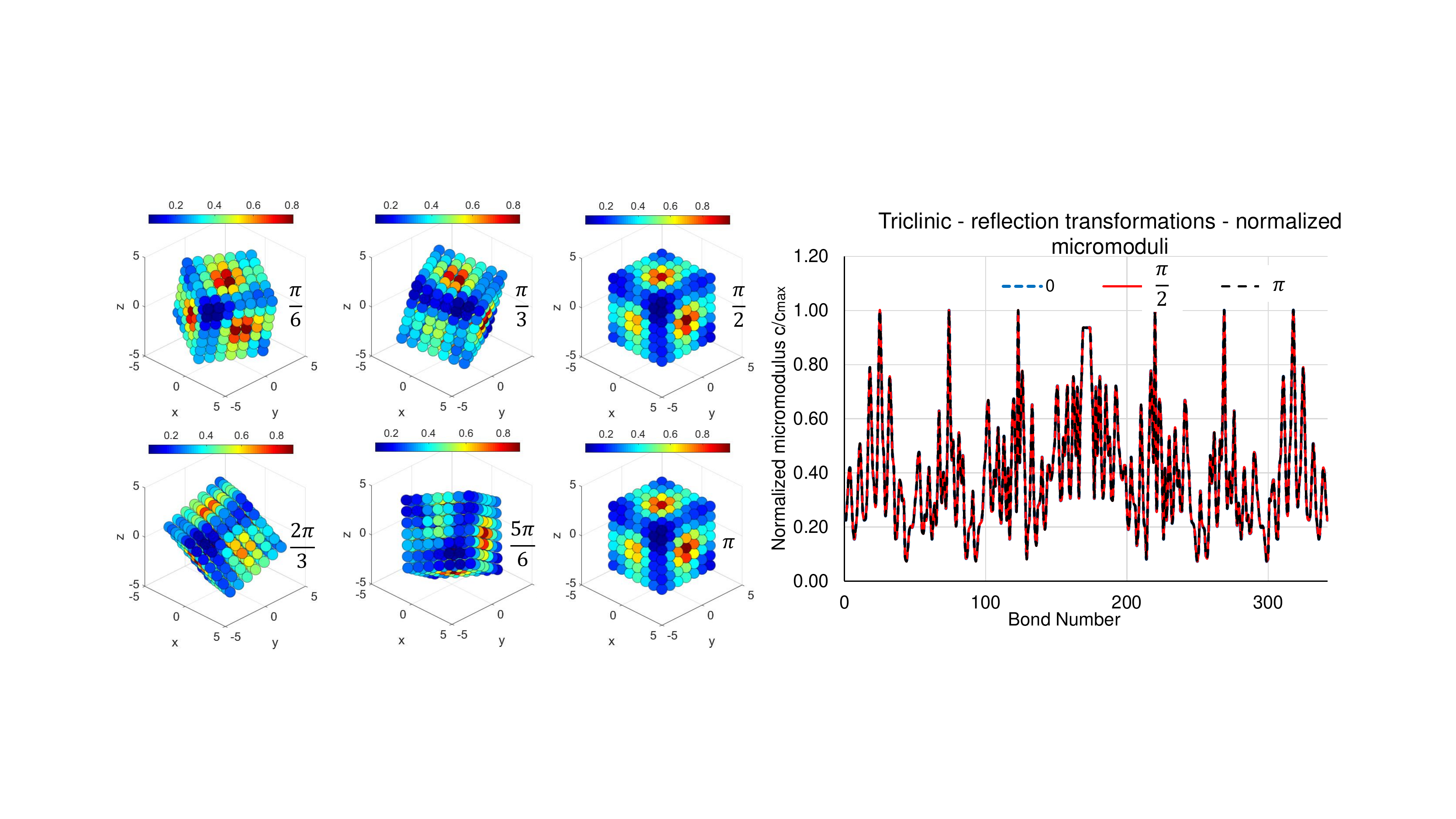}
	\caption{Rotation transformations of $\frac{\pi}{6}, \frac{\pi}{3}, \frac{\pi}{2}, \frac{2\pi}{3}, \frac{5\pi}{6}, \pi$ about all three axes (left) and distribution of bond micromoduli versus bond number (right) for triclinic symmetry.}
	\label{tric_rotation}
\end{figure}
For this example, six rotations of $\frac{\pi}{6}, \frac{\pi}{3}, \frac{\pi}{2}, \frac{2\pi}{3}, \frac{5\pi}{6}, \pi$ are chosen. These rotations are chosen arbitrarily and do not hold any special significance for this symmetry. Figure \ref{tric_rotation} shows the size cases of rotations (left) for which the overall error in the calibration is once again unchanged at 3.1873\%. Instead of plotting the normalized micromoduli for all cases, only those for $0, \frac{\pi}{2}, \pi$ are shown for brevity and to demonstrate that for these cases, the distribution of micromoduli are equal. It is quite obvious that a rotation of $\pi$ is equivalent to the original configuration. As it turns out, an equi-angle rotation of $\frac{\pi}{2}$ results in $\bm{R}$ being equal to the identity transformation $\bm{I}$ which indeed is within the symmetry group of triclinic materials. For all other rotations, the distribution of micromoduli are different from the original configuration and from each other. Therefore, for the triclinic symmetry out of all cases tested, both the conditions in Eq. \eqref{ver_3} and Eq. \eqref{ver_7} are satisfied only for the identity transformation, however the first condition is satisfied for all cases.
\begin{figure}[H]
	\centering
	\captionsetup{font=footnotesize}
	\captionsetup{justification=centering}
	\includegraphics[width=1\textwidth, trim={3cm 3.5cm 3cm 3.5cm},clip]{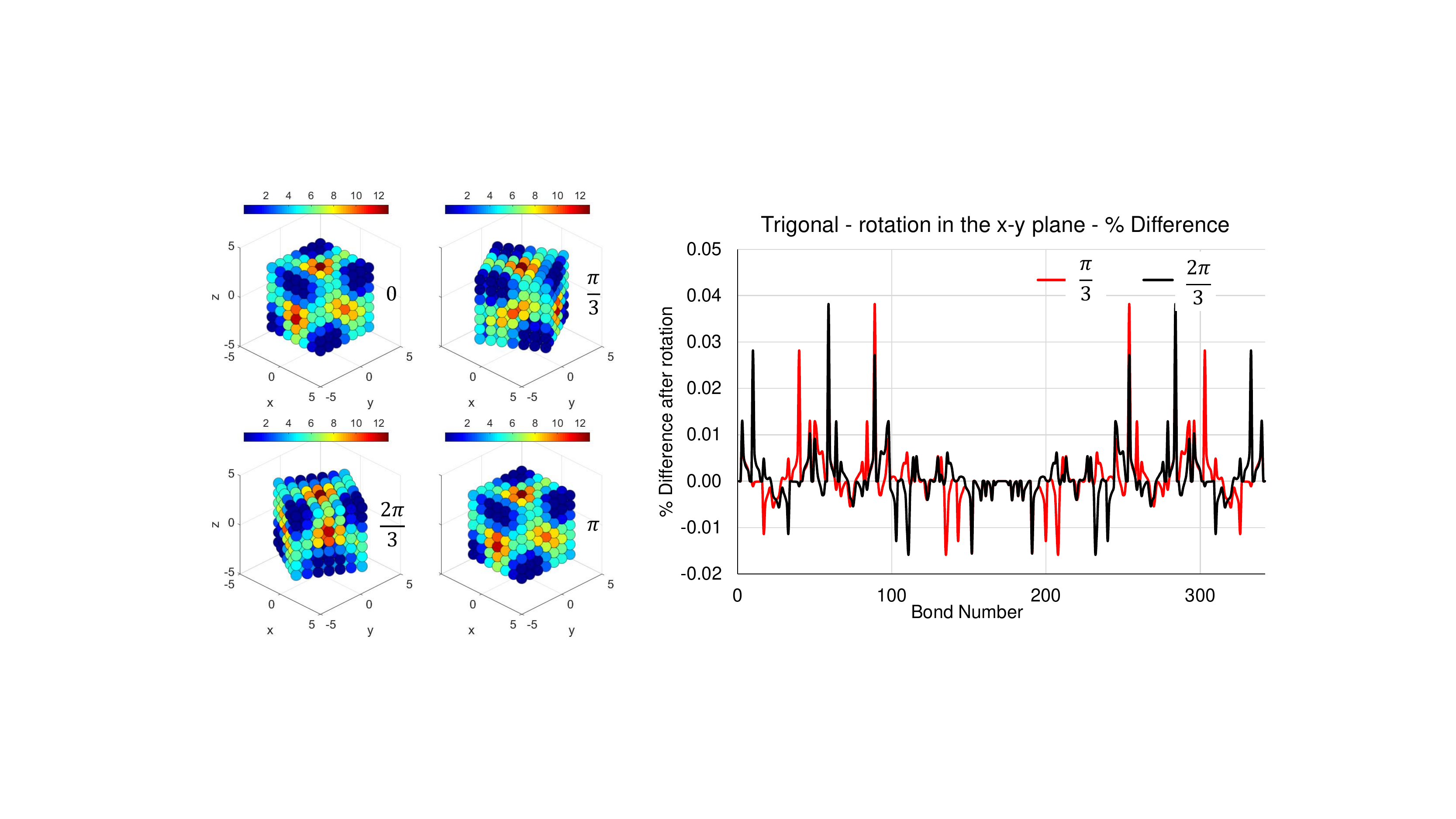}
	\caption{Rotation transformations of $\frac{\pi}{3}, \frac{2\pi}{3}$ and $\pi$ about all three axes (left) and \% difference in the bond micromoduli from the unrotated configuration versus bond number (right) for trigonal symmetry.}
	\label{tri_rotation}
\end{figure}

The second symmetry chosen is the trigonal material which has a few symmetry planes other than the ubiquitous identity and inversion transformations in it's symmetry group. Figure \ref{tri_rotation} shows the unrotated configuration along with three rotations of $\frac{\pi}{3}, \frac{2\pi}{3}$ and $\pi$ about the $\bm{e}_3$ axis, i.e. in the $x-y$ plane. In all cases, the effective peridynamic stiffness tensor is identical to the one found in section \ref{section_8_symm} with the error of calibration constant at 2.632\% thus satisfying the first condition Eq. \eqref{ver_3}. Unlike in the triclinic case where the equi-angle rotation of $\pi/2$ resulted in the cubic lattice being the same as the original configuration and therefore the micromoduli being identical, in the current case rotations of $\frac{\pi}{3}$ and $\frac{2\pi}{3}$ fall within the symmetry group of the trigonal material. Note that the lattice after rotations is not orthogonal to the coordinate system. The plot in Figure \ref{tri_rotation} (right) shows the \% difference in the micromoduli resulting from a rotation of $\frac{\pi}{3}$ and $\frac{2\pi}{3}$ from the original configuration. It is seen that the difference is not identically zero however is negligible. The maximum absolute \% difference is found to be $<0.04\%$ which explains the negligible impact on the calibration error. Thus the second condition given in Eq. \eqref{ver_7} can be said to satisfied as well. In other rotation transformations tested which did not fall under the symmetry group of the material (not presented here for the sake of brevity), only the first criteria given in Eq. \eqref{ver_3} was satisfied as expected.

\begin{figure}[H]
	\centering
	\captionsetup{font=footnotesize}
	\captionsetup{justification=centering}
	\includegraphics[width=1\textwidth, trim={2cm 3.5cm 2cm 3.5cm},clip]{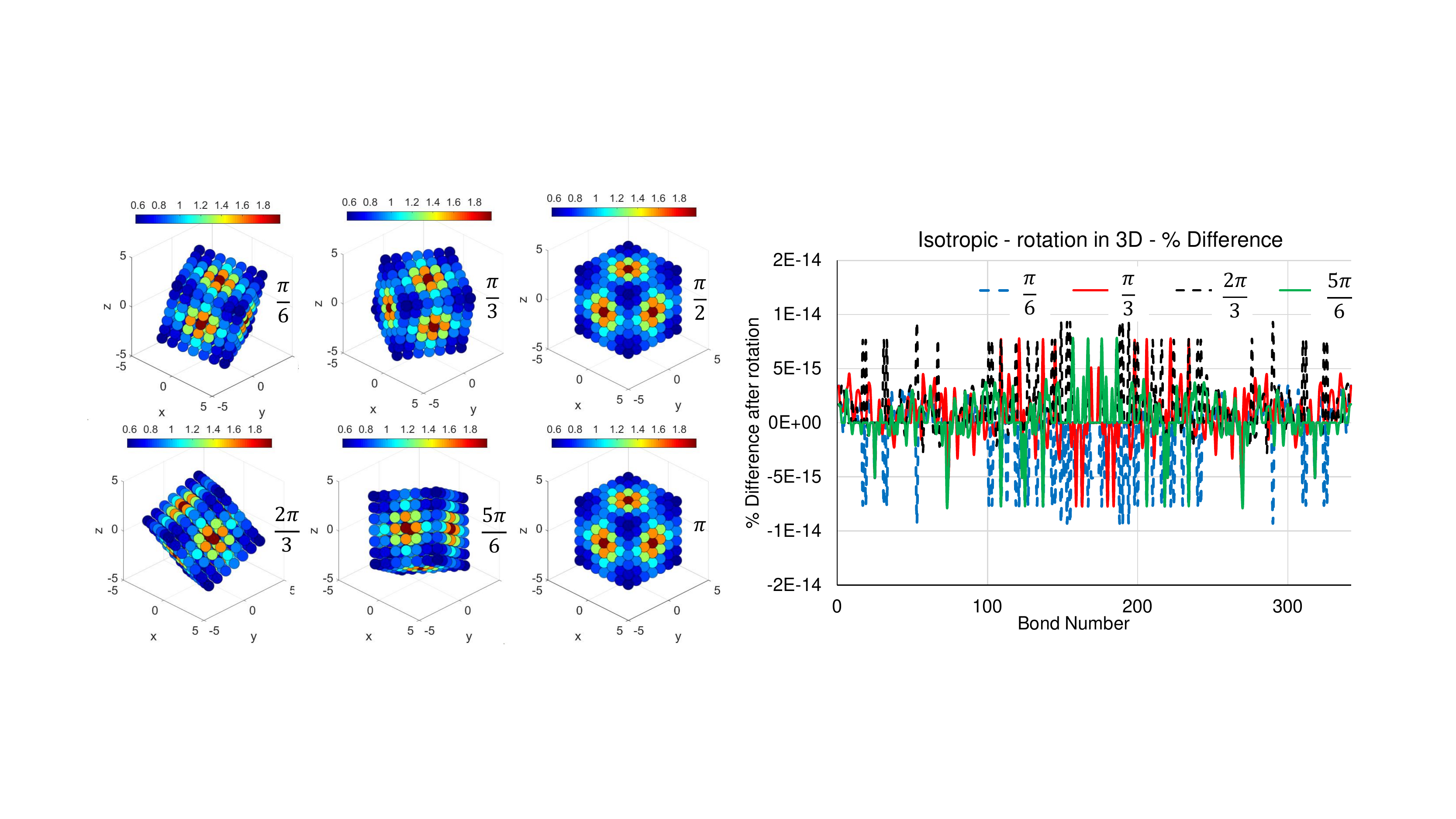}
	\caption{Rotation transformations of $\frac{\pi}{6}, \frac{\pi}{3}, \frac{\pi}{2}, \frac{2\pi}{3}, \frac{5\pi}{6}, \pi$ about all three axes (left) and \% difference in the bond micromoduli from the unrotated configuration versus bond number (right) for isotropic symmetry.}
	\label{iso_rotation}
\end{figure}
Lastly, isotropic symmetry, the least anisotropic with infinitely many transformations in it's symmetry group, is chosen to demonstrate the effect of lattice rotations. Similar to the triclinic case, equi-angle rotations of  of $\frac{\pi}{6}, \frac{\pi}{3}, \frac{\pi}{2}, \frac{2\pi}{3}, \frac{5\pi}{6}, \pi$ are chosen as shown in Figure \ref{iso_rotation}. As in the triclinic case, rotations of $\frac{\pi}{2}$ and $\pi$ result in a return to the original unrotated configuration and therefore quite obviously, the distribution of micromoduli is identical. Note however, that even with a visual inspection it is easy to see that the micromoduli for all cases of rotations are identical, no matter the rotation imposed. The plot shows the \% difference in the micromoduli from the original unrotated configuration which demonstrates that the change in micromoduli is basically zero (close to machine precision). In addition, since the material (Pyroceram 9608 in this case) exactly satisfies Cauchy's relations, the error in calibration is also found to be zero for all cases thus satisfying both conditions Eq. \eqref{ver_3} and Eq. \eqref{ver_7}.

\section{Summary and Conclusions}
A novel and general method to evaluate micromoduli for anisotropic bond-based models has been presented. Most literature has focused on specific models for certain material symmetries but little work has been done for incorporating anisotropy in bond-based peridynamics in a general way. The numerical method presented in this work is a first in that it can be used for any of the eight elastic material symmetry classes. The proposed method is based on a least-squares system of equations where the bond micromoduli are treated as the unknowns to be solved for, such that the error in the peridynamic stiffness tensor and the reference material stiffness tensor is minimized in a least-squares sense.

The peridynamic stiffness tensor, due to the limitation of bond-based peridynamics, inherits some conditions on the stiffness tensor known as Cauchy's relations along with the more traditional major and minor symmetries. The implications of Cauchy's relations for each material symmetry are presented and an example material is chosen for each symmetry to demonstrate the advantage of the proposed numerical method. It is shown that the resulting peridynamic stiffness tensor always satisfies Cauchy's relations, without any a priori modification. Cauchy's relations emerge naturally from the proposed numerical method of calibration. Therefore, if the reference material satisfies Cauchy's relations a priori, the error in calibration of the bond micromoduli is effectively zero. Since the proposed method works for a discretized neighborhood and does not need an analytically evaluated value of micromodulus, no further volume corrections are required.

Extensive parametric studies are conducted to demonstrate that the numerical method is general enough to cater to a large variety of horizon sizes, influence functions, neighborhood shapes and lattice rotations. It is found that if the lattice rotation falls under the symmetry group of the material, then not only is the peridynamic stiffness tensor unchanged from the original configuration, it is also seen that the rotated peridynamic stiffness tensor is invariant as well, thus serving as model verification. 

This work opens up many avenues for future development of both, fundamental aspects of anisotropic bond-based peridynamics as well as application to real-world problems. 

\section{Acknowledgements}
This work has been funded in part by Corning Incorporated. The author thanks Jason T. Harris and Ross. J. Stewart for discussions, suggestions and support. The author would also like to acknowledge the continued support for peridynamics from Sam S. Zoubi.

\bibliographystyle{unsrt}
\bibliography{References}

\appendix

\section{Universal Anisotropy Index} \label{UAI}

Assuming that $\bf{C}$ is the stiffness tensor written in Voigt notation, the universal anisotropy index can be evaluated as follows \cite{ranganathan2008universal, doi:10.1038/sdata.2015.9},

\begin{equation}
\bf{s} = \bf{C}^{-1},
\end{equation}
\noindent where $\bm{s}$  is the compliance matrix in Voigt notation. The Voigt and Reuss estimates for the bulk and shear modulus are,
\begin{equation}
9K_V = \left(C_{11} + C_{22} + C_{33}\right) + 2\left(C_{12} + C_{23} + C_{31}\right)
\end{equation}
\begin{equation}
1/K_R = \left(s_{11} + s_{22} + s_{33}\right) + 2\left(s_{12} + s_{23} + s_{31}\right)
\end{equation}
\begin{equation}
15G_V = \left(C_{11} + C_{22} + C_{33}\right) - \left(C_{12} + C_{23} + C_{31}\right) + 3\left(C_{44} + C_{55} + C_{66}\right)
\end{equation}
\begin{equation}
15/G_R = 4\left(s_{11} + s_{22} + s_{33}\right) - 4\left(s_{12} + s_{23} + s_{31}\right) + 3\left(s_{44} + s_{55} + s_{66}\right)
\end{equation}
\noindent The average of the Voigt and Reuss estimates are,
\begin{equation}
2K_{VRH} = (K_V + K_R)
\end{equation}
\begin{equation}
2G_{VRH} = (G_V + G_R)
\end{equation}
\noindent The universal anisotropy index is,
\begin{equation}
A^{U} = 5\frac{G_V}{G_R} + \frac{K_V}{K_R} - 6 \geq0
\end{equation}

\section{Isotropic Symmetry with Varying Micromoduli} \label{Isotropic symmetry}
While a non-constant micromodulus is counter-intuitive to the idea of an `isotropic' bond-based peridynamic material, we defer to the well-established and broader notion of  material symmetry of the \textit{elasticity tensor} rather than the isotropy of the pairwise force function \cite{silling2000}. In other words, we defer to idea that the effective peridynamic stiffness tensor should be invariant under all orthogonal transformations rather than the invariance of the pairwise force function itself. As mentioned previously in Eq. \eqref{Cijkl_inv}, one such transformation, namely the inversion transformation can easily be observed in the calibrated neighborhoods. The mirror images of bonds across the center of the neighborhood carry exactly the same micromodulus value. Even from a practical standpoint, the resulting deformation of a peridynamic model is a function of deformation of all bonds in the neighborhood of a particle and it is rare, if at all that a single bond is used to describe a material. Moreover, to model an isotropic material, the analytical formulation of micromoduli can be used instead of the numerical method described here. This method is however expected to be highly beneficial to other, more complex material symmetries.

\end{document}